\documentclass[]{fairmeta}
\usepackage{booktabs}
\usepackage{multirow}
\usepackage{graphicx}
\usepackage{hyperref}
\usepackage{xspace}
\usepackage{amsfonts}
\usepackage{wrapfig}
\usepackage{xcolor}
\usepackage{dsfont}
\definecolor{myblue}{HTML}{61CBF4}
\definecolor{mypink}{HTML}{E59EDD}
\definecolor{myred}{HTML}{D14A4A}
\definecolor{darkgreen}{RGB}{0,100,0}
\definecolor{burgundy}{RGB}{128,0,32}

\title {\textcolor{myblue}{Wispy} to \textcolor{myred}{Voluminous}: Prior-free Multi-view Capture of Strand-level Facial Hair}
\author[1,3,*]{Jaeseong Lee}
\author[2]{Giljoo Nam}
\author[3]{Adrian Jarabo}
\author[3,\dag]{Carlos Aliaga}
\affiliation[1]{KAIST}
\affiliation[2]{Meta Codec Avatar Lab}
\affiliation[3]{Meta Reality Labs Research}
\contribution[*]{Work done during an internship at Meta}
\contribution[\dag]{Project leader and corresponding author.}
\abstract{
Facial hair is a defining trait of personal identity, yet remains a critical bottleneck for digital avatars. Recent volumetric methods achieve photorealism but bake hair into the underlying face geometry, preventing editability and failing to resolve sparse, strand-like structures. Meanwhile, scalp-hair reconstruction methods target dense hair volumes and do not transfer to the sparse, spatially-varying nature of facial hair. We present a pipeline that automatically reconstructs facial hair---beard, mustache, lashes, and brows---from multi-view images, converting an unstructured 3D Gaussian representation into an explicit curve-based strand representation. We resolve geometric ambiguities in four stages: (i) optimizing 3D Gaussians constrained by tracked head geometry to enforce early ray termination and suppress sub-surface noise; (ii) tracing continuous strands robust to frequent crossings and extreme curvature; (iii) grounding strands to the surface and resolving root--tip ambiguity via a physically-motivated prior; and (iv) refining the reconstruction through opacity-driven density control under photometric optimization. To our knowledge, this is the first method to reconstruct high-fidelity facial hair strands from a 3D Gaussian representation. The recovered strands faithfully preserve the orientation and sparsity patterns characteristic of facial hair, and yield assets immediately suitable for downstream production tasks, including facial animation and physical simulation, geometric grooming and transfer, appearance editing, and physics-based rendering.
}\label{abstract}

\date{\today}
\correspondence{Jaeseong Lee at \email{wintermad1245@kaist.ac.kr}}

\metadata[Website]{\url{https://summertight.github.io/gaussiantrimmer}}
\begin{document}

\maketitle

\section{Introduction}
\label{intro}

Facial hair plays a central role in facial perception and identity. Eyebrows modulate expression, eyelashes are critical for realistic gaze, and beards and mustaches serve as defining personal appearance with strong cultural roots. Unlike scalp hair, facial hair is sparse, locally structured, and tightly coupled to facial anatomy, with frequent crossings and occlusions that amplify reconstruction ambiguity and leave little redundancy to mask errors, making strand-level accuracy essential. 

Current state-of-the-art facial modeling falls into two approaches: On one hand, mesh-based pipelines~\cite{bharadwaj2023flare,feng2021deca,lattas2023fitme,papantoniou2023relightify,retsinas2024smirk,liu2025bioface} focus on reconstruct facial geometry and neglect facial hair, or treat it as texture; this requires manual creation (grooming) of strand-based facial hair, even in high-quality assets. 
On the other hand, volumetric approaches~\cite{lee2025surfhead,qian2024gaussianavatars,giebenhain2023nphm,zheng2022imavatar,zheng2023pointavatar,gafni2021nerface,saito2024rgca,lombardi2019neuralvolume,lombardi2021mvp,lee2025texavatars} are capable to reconstruct both the face geometry and facial hair with impressive visual realism, but encode both as a continuous density field that entangles geometry, illumination, and material, making it poorly aligned with 3D content creation tools.

Our goal is to automatically reconstruct facial hair using explicit curves: Curve-based grooms are a natural way of modeling hair, and support rich editing, physical simulation, and tight integration with facial animation systems, backed by decades of tooling. 
Although several works have explored curve-based facial hair reconstruction~\cite{li2023ems,beelerfacialhair,kerbiriou2024eyelash}, the landscape remains fragmented and focused on specific subcategories (lashes, eye brows). Moreover, existing works fail at grounding strands to the facial geometry, and in the presence of complex crossings. Bridging multi-view image-based capture with the editability and physical interpretability of strand-based grooms remains an open challenge for production-ready facial hair reconstruction.
Furthermore, simply adapting dedicated scalp-hair reconstruction techniques to the face is non-trivial, as the two domains differ fundamentally in structure. Existing approaches~\cite{groomcap,gaussianhaircut,sklyarova2023neuralhaircut,wu2024monohair,rosu2022neuralstrands} are designed under the implicit assumption of a largely uniform, dense hair volume, and thus struggle with the sparse, spatially localized nature of facial hair, which exhibits spatially varying density and frequently reveals the underlying skin. Moreover, most state-of-the-art methods rely on learned priors trained on scalp hair, which do not generalize to facial hair. 

\begin{figure*}[t!]
\centering
 \includegraphics[width=\linewidth]{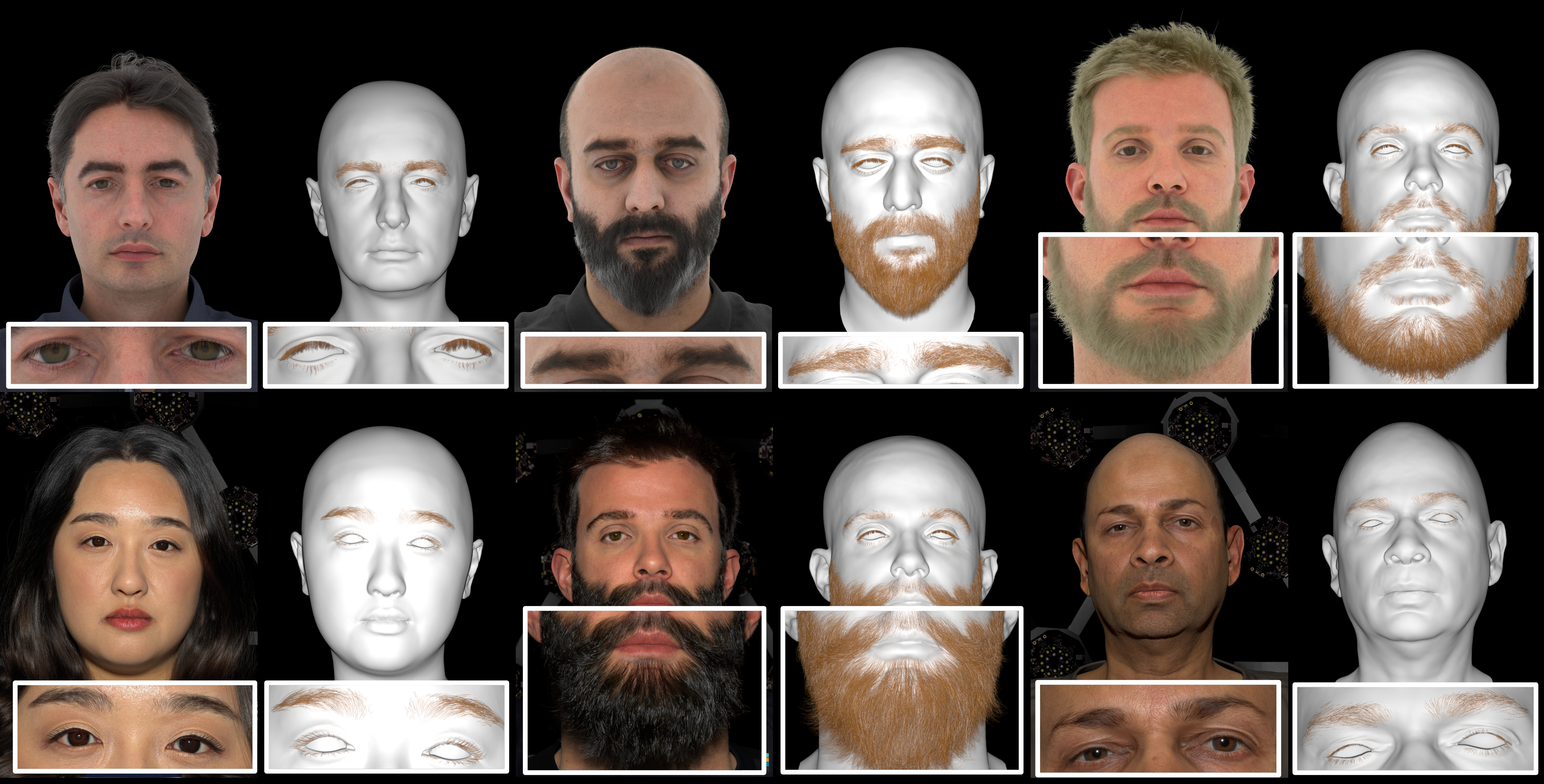}
   \caption{ Our pipeline reconstructs detailed hair geometry over a wide range of facial-hair types, including eyelashes, eyebrows, and beards. For each example, the input image is shown on the left, and a rendering of the reconstructed model from the corresponding viewpoint is shown on the right. }
    \label{fig:teaser}
\end{figure*}

In this work, we present a prior-free approach for reconstructing editable, strand-level facial hair geometry from multi-view images. Our method handles the full spectrum of facial hair types -- from dense beards to delicate eyelashes, see \Cref{fig:teaser} -- by leveraging 3D Gaussian splatting (3DGS)~\cite{kerbl20233dgs} to capture high-frequency orientation fields while introducing strong geometric constraints tailored to the sparse, discontinuous nature of facial hair.
Our method is built on four core technical innovations: 
\begin{itemize}
    \item A depth-gated optimization that integrates the reconstructed head mesh into the 3DGS optimization to resolve the ambiguity between hair and skin;
    \item A Gaussians-to-strands procedure using a crossing-aware strand tracing pipeline based on continuous Euler integration, which recovers continuous strands even under frequent crossings or extreme curvature;
    \item A physically-plausible strand grounding, that resolves the root-tip ambiguity using a region-adaptive strategy adapted to lashes, brows and beards, and that places strands roots to the head mesh;
    \item And a final adaptive strand surgery, that finetunes strands via a photometric refinement. 
\end{itemize}
Out method is fully automatic, and producing a compact, production-ready explicit facial-hair asset. 
This transition from implicit volumes to explicit geometry is what distinguishes our approach from pure view-synthesis methods. By converting the captured volume into structured strands, \textbf{we bridge the gap to traditional computer graphics pipelines, which allows editing, works with physics simulation, and can be rendered in any modern renderer}.

% \begin{teaserfigure}
%   \centering
%   \includegraphics[width=1.0\textwidth]{figures/mock_teaser_v3.png}
%   \caption{Our pipeline reconstructs detailed hair geometry over a wide range of facial-hair types, including eyelashes, eyebrows, and beards. For each example, the input image is shown on the left, and a rendering of the reconstructed model from the corresponding viewpoint is shown on the right. %\adrian{add closeups to the beards, the lashes and eyebrows. Not to the full eye region. And use different insets for different characters. E.g., Yaser should have the inset to the beard and moustache. Violet and Lambamba to the brows. Ronald lashes. Crop negative space on the top, and also at the bottom of the renders. Make sure the insets match pixel-wise between referene and reconstruction. s}
%   }
%   \label{fig:teaser}
% \end{teaserfigure}

\section{Related Work}
\label{sec:related}

\paragraph{Hair Reconstruction}
% \subsection{Hair Modeling}
Early hair reconstruction methods primarily relied on hand-crafted appearance and orientation cues: \citet{grabli02} formulated hair recovery as an inverse appearance problem based on pixel-wise reflectance variations, while \citet{paris04,paris08} extended this direction by estimating multi-view 2D/3D orientation fields, triangulating volumetric hair samples, and tracing explicit strands through the recovered volume using forward Euler integration. Subsequent multi-view stereo and line-based matching methods~\citep{luo2012multi, luo2013wide, lpmvs} further improved robustness under complex strand structures, while inverse rendering approaches~\citep{hairinverserendering, gaussianhair, groomlight} jointly optimized strand geometry and appearance under physically-based reflectance models. Other works explored specialized sensing modalities such as computed tomography \citep{ct2hair} and thermal imaging \citep{herrera2012lighting}, although their practical applicability is limited by capture constraints and hardware requirements.

More recent approaches introduce learned priors \citep{hu2015uschair} together with volumetric or Gaussian-based intermediate representations \citep{rosu2022neuralstrands, sklyarova2023neuralhaircut, gaussianhaircut, wu2024monohair}. 
In parallel, \citet{Chang2025IPHG} showed that reconstructed strands can be converted into compact, editable procedural grooms, highlighting the importance of downstream usability. Nevertheless, most existing pipelines remain fundamentally scalp-centric. 
\citet{groomcap} and \citet{takimoto2024drhair} assumed scalp-rooted strand organization and explicitly reconnect floating strands to a fitted head surface, which is appropriate for scalp hair but less suitable for facial hair, whose roots are spatially distributed across semantically distinct regions such as eyebrows, eyelashes, and beard. A parallel line of research addresses hair reconstruction from single images or text prompts \citep{rosu2025difflocks, sklyarova2025im2haircut, zheng2023hairstep, wu2022neuralhdhair, groomgen}, leveraging strand generative models trained on synthetic data to compensate for the ill-posed nature of monocular capture.

\paragraph{Facial Hair Reconstruction}
%
% \subsection{Facial Hair Modeling}
Early work on facial hair reconstruction focused on controlled multi-view capture setups. \citet{beelerfacialhair} proposed a coupled reconstruction framework that jointly models sparse facial hair fibers and the underlying skin surface, and \citet{winberg2022facial} extended this to dynamic performance capture, jointly modeling facial motion and hair deformation. 
More recent approaches explore strand-level modeling and learning-based techniques: \citet{li2024strandfacialhair} introduced a strand-accurate multi-view reconstruction pipeline leveraging line-based stereo cues \citep{lpmvs} and temporal optimization. 
Other approaches focus on specific regions of facial hair: 
\citet{li2023ems} reconstruct 3D eyebrow fibers from a single image, while \citet{kerbiriou2024eyelash} present a data-driven eyelash model from multi-view imagery enabling controllable editing, and \citet{xiao2021eyelashnet} focus on eyelash matting for downstream modeling.

Despite these advances, existing methods either rely on specialized capture setups, focus on isolated facial regions, or produce strands that are not grounded to the facial surface. Our work addresses these gaps by recovering editable, animation-ready strand assets directly from multi-view imagery, with explicit mesh grounding that ensures animation-readiness. As a direct consequence, per-frame hair tracking becomes unnecessary: Once grounded, strands can be driven purely by facial mesh deformation, enabling seamless integration with standard facial animation pipelines.

\begin{figure*}[t!]
\centering
    \includegraphics[width=\linewidth]{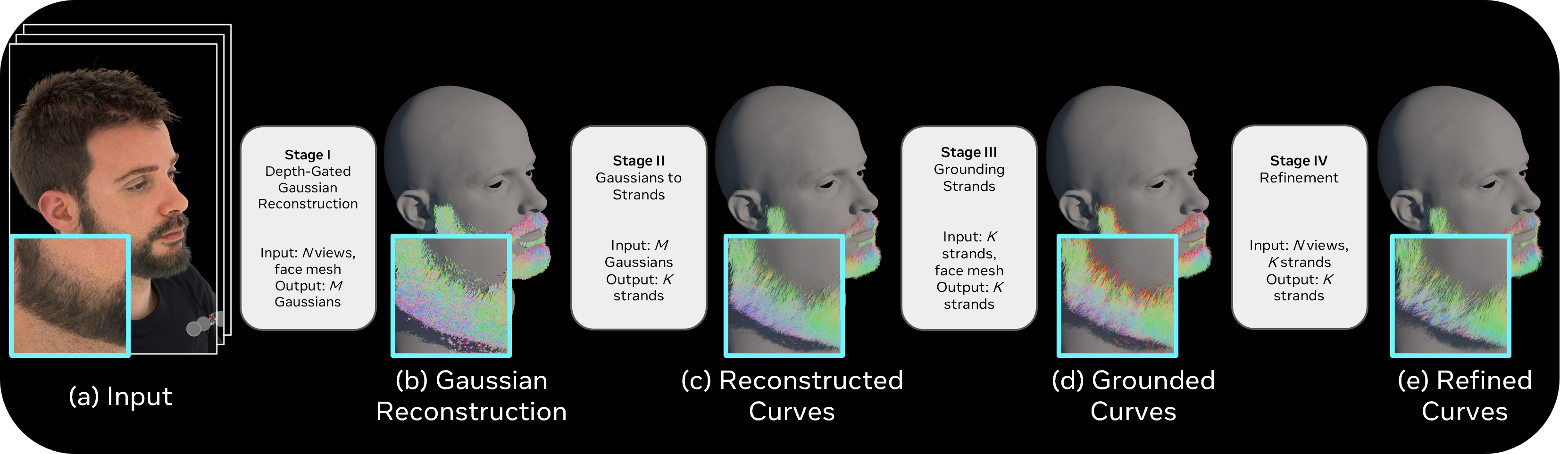}
   \caption{\textbf{Overview: }Our method reconstruct high-quality strand-based facial hair in four stages. From an input set of $N$ multi-view images (a), we first reconstruct a clean set of Gaussians (b), which drives our initial curve reconstruction (c). These are later grounded to the face mesh (d) to ensure they are compatible with facial animation and physical simulation. Finally, we refine our reconstruction by means of a photometric optimization (e), to ensure it matches the density and distribution of the input.
   }
    \label{fig:overview}
\end{figure*}

% \begin{figure*}[h]
%     \centering
%     \includegraphics[width=\linewidth]{figures/fig_pipeline_v4.png}
%     % \includegraphics[width=\linewidth]{figures/fig_pipeline_mockup_v3.png}
%     \caption{\textbf{Overview: }Our method reconstruct high-quality strand-based facial hair in four stages. From an input set of $N$ multi-view images (a), we first reconstruct a clean set of Gaussians (b), which drives our initial curve reconstruction (c). These are later grounded to the face mesh (d) to ensure they are compatible with facial animation and physical simulation. Finally, we refine our reconstruction by means of a photometric optimization (e), to ensure it matches the density and distribution of the input.
%     % \adrian{Figure and examples are great, but it takes lots of real state above and in the insets. I would crop a bit the top part of the fig on the left, to make it a bit more compact. Also, i would put the insets on top of the reconstruction, probably in the bottom left corner.  Done a quick mockup in Paint to illustrate what I mean: It is the same fig, but with some cropping and moving the insets.}
%     }
%     \label{fig:overview}
% \end{figure*}

\section{Facial Hair Capture}
Our pipeline reconstructs strand-level facial hair geometry in four stages. Our input is a set of multi-view images with uniform illumination, from which we extract a facial mesh using standard photometry. The output is the set of curves defining each hair, grounded to the facial mesh, with each strand's root position defined by the UV coordinate of the intersection point at the mesh. 

\Cref{fig:overview} shows an overview of our method. 
\textbf{Stage~I} optimizes a 3D Gaussians-based orientation field from 2D orientation cues by integrating a tracked mesh into the our new depth-gated rasterization process, which corrects the standard alpha-compositing behavior of 3DGS. 
\textbf{Stage~II} transforms the 3D Gaussians into structured facial-hair strands using a robust line-geometry processing algorithm that connects Gaussian primitives into continuous curves.
Next, \textbf{Stage~III} grounds the traced strands onto the tracked mesh while resolving the root-tip ambiguity using a physically-plausible grounding. This is one of the most challenging aspects of facial hair reconstruction, and unlike scalp hair, where strands share coherent directional neighborhoods over large regions, facial hair exhibits highly localized coherence. 
%
% Unlike scalp hair, where only the root lies near the surface, facial hair strands often have both endpoints in close proximity to the mesh, making it difficult to determine the growing direction. It also grows strands that lie far from the mesh to capture detached or floating structures, making this stage a key step toward animation-ready assets.
% \adrian{Wait, these are four stages, and before we said there were only three! Also, stages are not subsections, I would use "Stage IV (Sec.\ref{sec:refinement})"} 
Finally, \textbf{Stage~IV} adaptively refines the reconstructed strands in a sparsity-aware manner, pruning, cutting, and redistributing them under multi-view supervision with our depth-gated rasterizer. This adaptive strand-surgery stage is particularly important for facial hair, whose sparse structure makes the reconstruction prone to spurious primitives bleeding into the underlying skin, and thus serves to suppress such artifacts while preserving plausible strand density.

%The resulting output is a set of physically plausible 3D strands, suitable for downstream applications such as geometry editing and appearance matching. Our extracted strands can be physically simulated, geometrically edited (e.g., trimmed, implanted or groomed), and relit under novel illumination by modifying hair material (BSDF) properties, all without the artifacts inherent to baked radiance fields. 

% \adrian{A fig here? Also, there is no direct mapping between these three stages and the four subsections. }
% \jesse{Which kind of pic would be good? :) Like pipeline overview??}

%\adrian{Done until here}
%\adrian{Copy in Sec.3.3: making normal-based root disambiguation~\cite{ct2hair} unreliable — strands are frequently grounded in the flipped orientation, producing physically implausible growth directions such as strands appearing to grow upward or directly into the skin. Backward growing strategies~\cite{groomcap} similarly struggle, as the narrow local consistency regions of facial hair cause premature termination or misdirected growth. Physically plausible grounding is therefore essential, not only for visual fidelity but also for downstream editing and animation compatibility.}

\subsection{Stage~I: 3D Gaussian Orientation Field via Depth-Gated Optimization} 
%\subsection{Stage~I. 3D Orientation Field Lifting via Depth-Gated Gaussian Optimization} 
\label{sec:depth_gating}

Following prior work~\citep{gaussianhaircut}, we lift a per-view 2D orientation field computed using Gabor filters~\cite{paris04} from calibrated multi-view images, and use it to supervise the optimization of 3D Gaussian orientations, with their orientation given by the principal axis of the covariance matrix. We adopt 3DGS \citep{kerbl20233dgs} as our underlying representation, as discrete primitives better capture the sparse, high-frequency strand--skin boundaries of facial hair compared to continuous volumetric fields.

\newcommand{\Real}{\mathbb{R}}
Similar to \citet{gaussianhaircut}, we use a standard 3DGS to rasterize per-pixel quantities $\mathbf{Q}(p)$ via alpha compositing,
\begin{equation}
\mathbf{Q}(p)=\sum_i T_i(p)\cdot \alpha_i(p)\cdot \mathbf{q}_i(p),
\qquad
T_i(p)=\prod_{j<i}\bigl(1-\alpha_j(p)\bigr),
\label{eq:3dgs}
\end{equation}
with $\alpha_i(p)\in \Real$ and $\mathbf{q}_i(p) \in \Real^7$ the transparency and features rendered by Gaussian $i$, including the Gaussian color and orientation, and a binary hair-mask. This results in a final AOV image $\mathbf{Q}(p)$ with color $\hat{\mathbf{C}}(p)$, hair mask $\hat{M}_{\mathrm{hair}}(p)$, and orientation $\hat{\mathbf{O}}(p)$.

% \begin{figure}[t]
%     \centering
%     \includegraphics[width=1.0\columnwidth]{Figures/Exp/fig_dgr_gaussians_v2.pdf}
%     \caption{\textbf{Effect of depth gating in Stage I.}
% Depth-gating allows to ensure that Gaussians solely represent facial hair, which results into better organized primitives. In contrast, previous approaches \cite{gaussianhaircut} are noisier near the skin, which difficults extracting clear strands. %\adrian{I don´t understand this last sentence: "Note that both of them are pre-filtered by alpha and (or) label value of Gaussians, each."}\jesse{It is outdated now I will remove}
% }
%     \label{fig:onlyhairgaussians}
% \end{figure}

Unfortunately, as shown in Fig.~\ref{fig:onlyhairgaussians} (right), this formulation leds the optimizer to create semi-transparent Gaussians across the subsurface region around the facial-hair area, using diffuse opacity accumulation to explain hair pixels. This not only increases memory and optimization cost, but more critically creates noise for later using these Gaussians for guiding the explicit strands. 

We solve this problem by leveraging the reconstructed tracked face mesh to obtain a per-pixel depth map $\mathcal{D}_{\text{mesh}}(p)$, rasterized via~\citet{Laine2020diffrast}, and replace $\alpha_i$ with a gated opacity
\begin{equation*}
    \textcolor{blue}{\tilde{\alpha}_i(p)}
    =
    \mathds{1}\!\left[d_i(p)\leq \mathcal{D}_{\text{mesh}}(p)+\varepsilon\right] \cdot \alpha_i(p),
\end{equation*}
where $d_i(p)$ is the depth of the $i$-th primitive and $\varepsilon=3\,\mathrm{mm}$ is a small tolerance set empirically to account for mesh registration error and hair root proximity.
Substituting $\textcolor{blue}{\tilde{\alpha}_i}$ for $\alpha_i$ in \Cref{eq:3dgs} gives the gated compositing rule
%\begin{equation*}
%\mathbf{Q}(p)=\sum_i \textcolor{blue}{\tilde{T}_i(p)}\,\textcolor{blue}{\tilde{\alpha}_i(p)}\,\mathbf{q}_i(p),
%\qquad
%\textcolor{blue}{\tilde{T}_i(p)}=\prod_{j<i}\bigl(1-\textcolor{blue}{\tilde{\alpha}_j(p)}\bigr),
%\end{equation*}
%
which removes contributions from primitives behind the tracked surface and prevents gradients from flowing to spurious sub-surface geometry.

\paragraph{Optimization.}
Following \citet{gaussianhaircut}, we use the standard 3DGS photometric loss
\(\mathcal{L}_{\mathrm{rgb}} = (1-\lambda)\,\mathcal{L}_{1} + \lambda\,\mathcal{L}_{\mathrm{D\text{-}SSIM}}\)
(with \(\lambda = 0.2\)) together with an orientation alignment loss \(\mathcal{L}_{\mathrm{dir}}\).
We also use mask supervision, directly computed from the gated foreground opacity $\mathcal{L}_{\mathrm{mask}} = \big\| M_{\mathrm{hair}} - \hat{M}_{\mathrm{hair}} \big\|_{1}$,
where $\|\cdot\|_1$ denotes the per-pixel mean absolute error.
%
% \adrian{Is it the sum or the mean?}.
%
The three terms are $\lambda$-blended together, giving a loss for Stage I defined as
\begin{equation}
    \mathcal{L_{\text{I}}}
    =
    \mathcal{L}_{\mathrm{rgb}}
    +\lambda_{\mathrm{dir}}\cdot\mathcal{L}_{\mathrm{dir}}
    +\lambda_{\mathrm{mask}}\cdot\mathcal{L}_{\mathrm{mask}},
\label{eq:loss_gs}
\end{equation}
where $\lambda_{\mathrm{dir}} = 0.1$ and $\lambda_{\mathrm{mask}} = 1.0$, respectively.

%%%%%%%%%%%%%%%%%%%%%%%%%%%%%%%%%%%%%%%%%%%%%%%%%%%%%%
%%%%%%%%%%%%%%%%%%%%%%%%%%%%%%%%%%%%%%%%%%%%%%%%%%%%%%
%%%%%%%%%%%%%%%%%%%%%%%%%%%%%%%%%%%%%%%%%%%%%%%%%%%%%%

\subsection{Stage~II: From Gaussians to Strands}
\label{sec:tracing}

Once we have a relatively clean 3D Gaussian field, in this second step we move to continuous strands that will serve the basis for the remaining steps of our pipeline. For that, we perform three main steps: We first further filter the Gaussians obtained in Stage I; then we build a set of strands based on the reconstructed Gaussians by Euler tracing the orientation field; and at the end we stitch strands that can be explained with a single longer strand. 

\paragraph*{Step II.1: Filtering}
Even with mesh-depth-gated training that suppresses sub-surface Gaussians, the reconstructed facial-hair field still contains nontrivial noise due to speculars and view-averaging artifacts, which results in spurious primitives. We therefore first apply a \textbf{filtering} step: we keep Gaussians with opacity $\alpha$ above a predefined threshold and near-needle anisotropy, measured by effective rank (erank)~\cite{erank} following \citet{erankGaussian}. In addition, we prune Gaussians whose center $\mu$ lies inside the tracked mesh to prevent penetration. Details can be found in \Cref{sec:stageii_denoising}. 

This gives us a set of clean Gaussian $\tilde{\mathcal{G}}=\{(\tilde\mu_i,\tilde v_i,\alpha_i)\}$, with $\tilde\mu_i$ their mean and $\tilde v_i$ its principal axis direction. Finally, we split each Gaussian along its principal axis into smaller primitives of roughly 0.1 mm. 
Since the Gaussian needles have heterogeneous principal-axis lengths, we uniformly resample each one along its principal axis with $0.1\,\text{mm}$ spacing, copying the direction and opacity from the parent, to prevent tracing disconnects caused by large inter-Gaussian gaps.
% ###
% \adrian{NEW Q: What opacity? How do you define the principal axis length? In terms of how many sigmas? What is the subdivided sigma of each Gaussian? I.e., if an input Gaussian has parameters $\mu, v, \sigma, \alpha$, I assume it will create $N=floor(2*l(\sigma)/0.1mm)+1$ Gaussians, such that Gaussian $i$ has parameters $\mu+i*0.1*v, v, f_0(\sigma), f_1(\alpha)$, with $f_0$ and $f_1$ functions that map the previous $\sigma$ and $\alpha$ to the new ones.}
% \jesse{Each Gaussian $(\mu_i, v_i, \sigma_i, \alpha_i)$ is split along its principal axis into $N = \lfloor \sigma_{\max} / 0.1\,\text{mm} \rfloor + 1$ smaller primitives, where $\sigma_{\max}$ is the principal axis scale (i.e., $1\sigma$ along $v_i$). The $j$-th split primitive has parameters
% \begin{equation}
% \left(\mu_i + j \cdot 0.1\,\text{mm} \cdot v_i,\;\; v_i,\;\; (\sigma_{\perp}, 0.1\,\text{mm}),\;\; \alpha_i\right),
% \end{equation}
% where $\sigma_{\perp}$ denotes the original transverse scales (copied unchanged) and the principal axis scale is fixed to $0.1\,\text{mm}$. Opacity is copied directly from the parent Gaussian.}
% ###
\paragraph{Step II.2: Tracing}
With the clean Gaussians-based orientation field, we now extract the strands. We build upon previous Euler-based tracing pipelines \citep{lpmvs, ct2hair, groomcap}, which treat the input as a directional point-set and recover strands by local forward integration. 

We set all Gaussians as potential seeds for strands using an opacity-based priority queue. For each strand, we extract a Gaussian $k$, and start two Euler paths, with initial point $\mathbf{p}_0 = \tilde{\mu}_k$ and direction $\mathbf{v}_0=\pm \tilde{v}_k$, by doing small steps of $\Delta s=0.3$ mm. For each step $n>0$, we get a candidate point $\tilde{\mathbf{p}}_{n}=\mathbf{p}_{n-1}+s\cdot\mathbf{v}_{n-1}$ and search for a neighbor Gaussians set $\mathcal{N}_{n}$ around that point, enforcing locality and similar orientation. If the set $\mathcal{N}_{n}$ is empty, the Euler path is terminated, returning a strand with $n-1$ vertices. Otherwise, we create a new vertex with position $\mathbf{p}_n$ and orientation $\mathbf{v}_n$, defined as an opacity weighted average of the Gaussians in $\mathcal{N}_{n}$ following
\begin{equation}
\mathbf{p}_{n}=
\frac{\sum_{k\in\mathcal{N}_{n}}\alpha_k\cdot\tilde\mu_k}
{\sum_{k\in\mathcal{N}_{n}}\alpha_k},
\qquad
\mathbf{v}_{n}=
\mathrm{normalize}\!\left(
\sum_{k\in\mathcal{N}_{n}}\alpha_k\cdot\tilde v_k
\right).
\label{eq:alpha_weighted_update}
\end{equation}
To avoid that Gaussians contribute to multiple strands, once we accept vertex $\mathbf{p}_n$ we remove Gaussians from the priority queue based on proximity to the segment $\mathbf{s}_{n-1}=(\mathbf{p}_{n-1},\mathbf{p}_{n})$. To avoid potential overlooks from previous approaches \citep{lpmvs,ct2hair,hairinverserendering}, we do not remove Gaussians if the angle of their orientation to the segment's orientation is higher than a threshold and they might result in a valid strand. This allows us to keep crosses, while effectively removing duplicated and spurious Gaussians. 
The final output of this tracing is a set of traced strands
$\Gamma=\{\gamma_1,\gamma_2,\ldots,\gamma_{N_s}\}$ with $N_s$ strands,
where each strand $\gamma_i\in\mathbb{R}^{N_k\times 3}$ denotes a 3D polyline with $N_k$ points.
More details can be found in \Cref{sec:tracing_app}.

\paragraph{Step II.3: Stitching}
The traced strand set $\Gamma=\{\gamma_i\}_{i=1}^{N_s}$ 
is often fragmented in practice, due to weak observations under view-dependent effects, primitive removal during filtering, and premature tracing termination. Nearby strands also frequently cross or partially overlap, making naive proximity-based stitching prone to zigzag artifacts. Inspired by \citet{beelerfacialhair}, we apply a three-stage stitching procedure, we execute sequentially and repeated regressively for 30 iterations until convergence:
\begin{equation*}
\Gamma^{\star}=\texttt{Stitch}(\Gamma)=\texttt{OverlapMerge}(\texttt{TipLink}(\texttt{EnclosedPrune}(\Gamma))),
\end{equation*}
where \texttt{EnclosedPrune()} removes strands almost entirely contained within a longer strand, verified by geometric proximity, tangent consistency, and projection monotonicity. \texttt{TipLink()} connects compatible endpoints under one-to-one matching, with crossing-aware junction cleanup to suppress kinks and clip-like artifacts. Finally, \texttt{OverlapMerge()} fuses strand pairs that partially trace the same physical structure, merging non-overlapping portions and discarding fully redundant strands.

\subsection{Stage~III. Grounding Strands to Roots}
\label{sec:grounding}
In the grounding stage, we anchor traced strands to the tracked face mesh surface. This step is essential for simulation readiness and downstream editability, e.g., trimming, UV-based grooming (re-planting), and transferring beard geometry across identities. A prerequisite for grounding is to resolve the intrinsic \emph{root--tip ambiguity} of traced curves. We address this with category-specific heuristics for eyelashes, eyebrows, and beard. 
% \adrian{Show an image?}

\subsubsection{Beard and mustache}
Prior hair reconstruction pipelines \citep{groomcap,ct2hair} typically assume that hair strands emerge approximately along the surface normal, accepting or rejecting strands based on normal consistency. However, for facial hair this assumption does not hold since: (i) surface-attached and volumetric components frequently coexist within the same region (e.g., skin-following stubble together with detached beard volume), and (ii) beard strands tend to follow low-angle, skin-aligned growth patterns with strong regional variation. Backward growing strategies~\cite{groomcap}, which grow the strands from surface to outward volume, similarly struggle, as the narrow local consistency regions of facial hair cause premature termination or misdirected growth. %Physically plausible grounding is therefore essential, not only for visual fidelity but also for downstream editing and animation compatibility.

We observed that beard growth direction is strongly correlated with gravity — strands tend to grow downward and outward from the skin surface rather than along the surface normal. This observation is consistent with clinical and cosmetic studies on beard growth patterns \citep{maurer2016malebeard, leao2017beard}, and empirically. %, using a gravity-blended reference direction substantially outperforms surface-normal-based disambiguation in our ablation (see Fig.~\ref{fig:ablation_grounding}). 
We therefore define a mixed reference direction
\begin{equation}
\mathbf{r}=\mathrm{norm}\big((1-\lambda_\mathbf{r})\cdot\mathbf{n}+\lambda_\mathbf{r}\cdot\mathbf{g}\big),
\label{eq:ref_dir}
\end{equation}
% where $\mathbf{n}$ is the tracked surface normal and $\mathbf{g}$ is the gravity direction \adrian{what is the value of $\lambda_\mathbf{r}$?}\jesse{actually, 0.7!}, and use $\mathbf{r}$ as the basis for root--tip disambiguation and grounding throughout the following steps.
where $\mathbf{n}$ is the tracked surface normal, $\mathbf{g}$ is the gravity direction, and $\lambda{=}0.7$ is empirically set to weight gravity over the surface normal, reflecting the dominant downward growth pattern of beard hair.

\paragraph{Step III.1: Surface consistency.}
We evaluate each traced strand endpoint against the reference direction $\mathbf{r}$~(Eq.~\ref{eq:ref_dir}). An endpoint is considered \emph{surface-consistent} if it satisfies two conditions: (a) its distance to the tracked mesh is within $\tau_c$, and (b) its strand direction deviates from $\mathbf{r}$ by less than $30^\circ$. If one endpoint passes it is assigned as the root and snapped to the surface. If both pass, the endpoint with higher cosine similarity to $\mathbf{r}$ is selected as the root, with distance serving as a tie-breaker. This step resolves root--tip ambiguity for the majority of beard strands close to the surface.

\paragraph{Step III.2: Voting-based disambiguation.}
Strands that remain ambiguous after Step~III.1 are resolved by a voting procedure over trusted grounded strands $\Gamma_t$. For each ambiguous strand $\gamma_c$, we query all trusted strands within a spatial radius. Each trusted strand $\gamma_t \in \Gamma_t(\gamma_c)$ casts a vote based on the sign of the cosine between its root-to-tip direction and the candidate strand direction: a positive cosine ($<90^\circ$) votes for one root hypothesis, and a non-positive cosine ($\geq 90^\circ$) votes for the other. If more than $60\%$ of neighbors support one hypothesis — a threshold set empirically — that endpoint is assigned as the root. Otherwise, the root is determined by a geometric fallback using the face normal and gravity direction. All strands are assigned a root at this stage.

\paragraph{Step III.3: Completion of flying strands.}
Strands whose root remains far from the mesh after Step~III.2 are completed by parallel translation toward the surface. Among trusted strands within a search radius, we select the one whose tangent direction is most aligned with the candidate strand (cosine similarity within $90^\circ$). We then follow this strand's step vectors in reverse, accumulating displacement until the root reaches the surface or the maximum grow distance $d_{\max}$ is exceeded. Strands that fail due to no valid neighbor or exceeding $d_{\max}$ are discarded.

% \paragraph{Step III.2: Voting-based disambiguation.}
% When strands lie near multiple surface regions, the surface-consistency produces ambiguous results. 
% %
% To disambiguate them, we build a set of \emph{trusted grounded strands} from the surface-consistency condition. For each candidate ambiguous strand $\gamma_c$, we query nearby trusted strands $\mathcal{S}_t(\mathbf{s}_c)$, and measure the orientation agreement between each $\mathbf{s}_t \in \mathcal{S}_t(\mathbf{s}_c)$ and the candidate $\mathbf{s}_c$ \adrian{How? What is the metric?}. If more than $60\%$ of neighbors — a threshold set empirically — support one root hypothesis, the corresponding endpoint is selected as the root; if it lies within $\tau_c$ to the tracked mesh, it is snapped to the surface.

% \paragraph{\textbf{Eyelash.}}
% Eyelash grounding is unambiguous: for each eyelash strand, we assign as the root the endpoint nearest to the tracked mesh and snap it to the surface.
\subsubsection{Eyelashes}
Eyelashes follow a more surface--normal-aligned growth pattern than the rest of facial hair, and their strands tend to remain in close proximity to the eyelid surface. We therefore adopt a simple grounding procedure for eyelashes: We ground strands purely based on proximity to the tracked mesh. Strands whose nearest endpoint lies within the distance threshold are directly snapped to the surface, and remaining flying strands are resolved through the same voting and completion procedure described above.

\begin{figure}[t]
    \centering
    \begin{subfigure}[t]{0.54\linewidth}
        \centering
        \includegraphics[width=\linewidth]{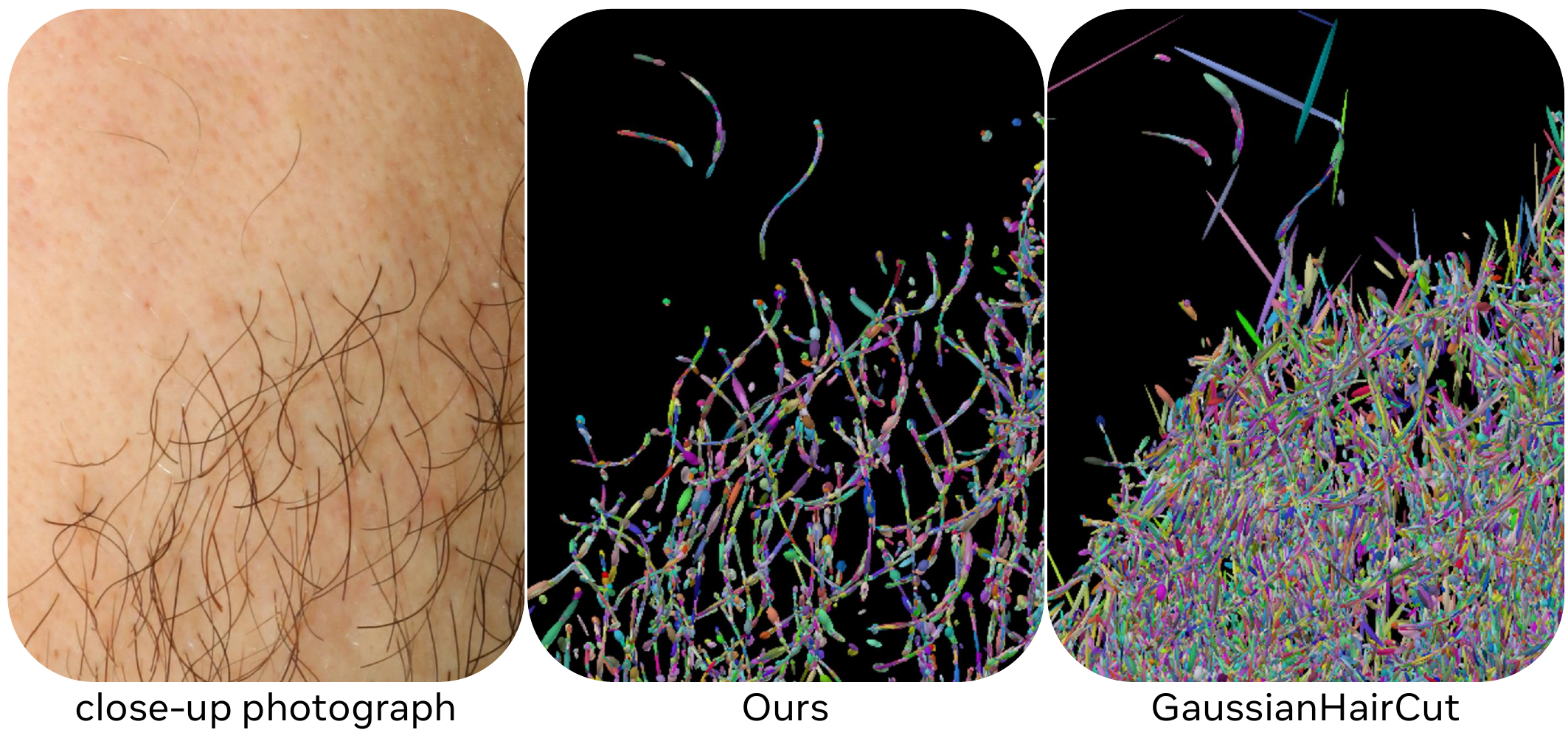}
        \caption{Effect of depth gating in Stage I.}
        \label{fig:onlyhairgaussians}
    \end{subfigure}
    \hfill
    \begin{subfigure}[t]{0.45\linewidth}
        \centering
        \raisebox{7.5pt}{\includegraphics[width=\linewidth]{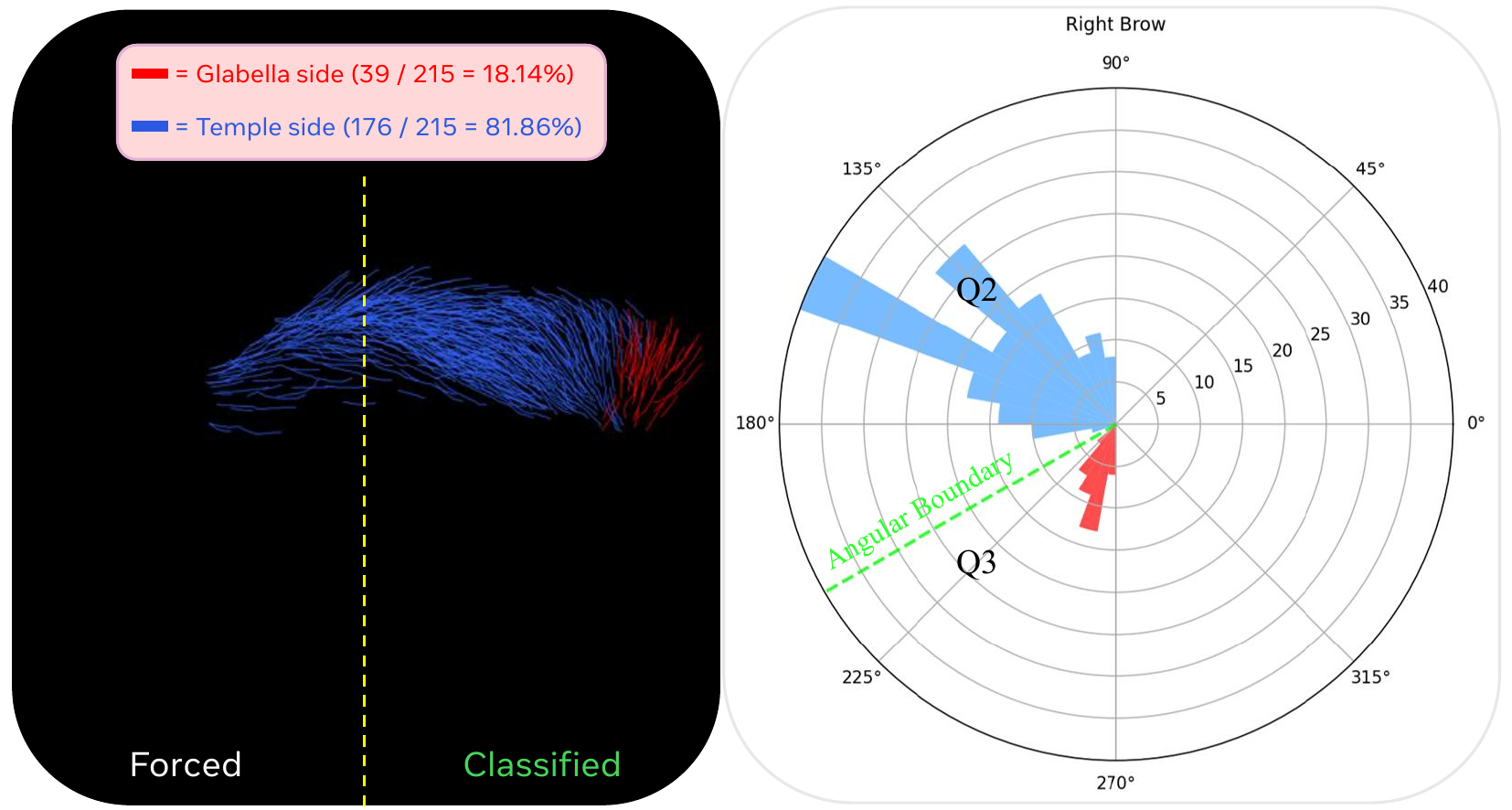}}
        \caption{Root-tip ambiguity resolve for eyebrows.}
        \label{fig:eyebrow_angular_dist}
    \end{subfigure}
    \caption{\textbf{(a)} Depth-gating ensures that Gaussians solely represent facial hair, yielding better organized primitives. Previous approaches~\cite{gaussianhaircut} are noisier near the skin, which makes extracting clear strands difficult.
    \textbf{(b)} Rose diagram of strand growth directions for the inner half of each eyebrow. Vectors are computed from the \textcolor{red}{glabella-side (inner)} to \textcolor{blue}{temple-side (outer)} endpoint in frontal projection. Strands below the angular threshold \textcolor{green}{(green, $30^\circ$)} are classified as root=outer, indicating ambiguous growth direction near the glabella. The angular distribution exhibits a \textbf{bimodal} pattern with a natural gap near the decision boundary ($\theta = 30^\circ$), validating the threshold-based classification.
    }
    \label{fig:combined}
\end{figure}

% \begin{figure}[t]
%     \centering
%     \includegraphics[width=0.99\linewidth]{Figures/Exp/fig_eyebrow_angular_v2.pdf}
%     \caption{\textbf{Visualization of root-tip ambiguity resolve process of eyebrows.} Rose diagram of strand growth directions for the inner half of each eyebrow. Vectors are computed from the \textcolor{red}{glabella-side (inner)} to \textcolor{blue}{temple-side (outer)} endpoint in frontal projection. Strands below the angular threshold \textcolor{green}{(green, $30^\circ$)} are classified as root=outer, indicating ambiguous growth direction near the glabella. The angular distribution exhibits a \textbf{bimodal} pattern, with a natural gap near the decision boundary ($\theta = 30^\circ$), validating the threshold-based classification}
%     \label{fig:eyebrow_angular_dist}
% \end{figure}

\subsubsection{Eyebrows}
We disambiguate eyebrow direction using a side-aware rule in mirrored frontal view. Consider the \emph{right eyebrow} in 3D space (which appears on the left side in screen space after mirroring). We project each strand to the front view and define an \emph{inner} endpoint (closer to the glabella) and an \emph{outer} endpoint (closer to the temple), and use a two-fold categorization, as described in \Cref{fig:eyebrow_angular_dist}:
For strands in the outermost eyebrow strip (among four evenly sliced vertical strips in screen space), the root assignment is self-evident: the inner endpoint must be the root. 
The only ambiguous region is the second strip. There, we translate each strand such that the inner endpoint is at the 2D origin and inspect the tip location. If the tip falls in Quadrant II, we keep the inner endpoint as the root. If it falls in Quadrant III (e.g., a strand whose tip bends toward the glabella), we switch the root to the outer endpoint \emph{only if} the strand direction deviates from the $x$-axis by more than $30^\circ$; otherwise we keep the inner root. We apply the same logic to the left eyebrow by mirror symmetry. After root assignment, we snap roots to the nearest tracked surface points.

\subsection{Stage~IV. Refinement Stage for Sparsity Matching}
\label{sec:refinement}
Even with our crossing-aware tracing, over-reconstruction remains a recurring failure mode. Tracing produces redundant nearly-parallel strands wherever local Gaussian support exists, and transmittance-based gradient attenuation leaves inner Gaussian primitives unpruned — both compounding over-dense reconstructions that neither grounding nor stitching is designed to correct. This is particularly problematic for facial hair because \emph{sparsity itself is part of the signal}. We therefore add a final refinement stage that explicitly aligns strand density with the \emph{observed} images.

\paragraph{Representation.}
Following \citep{gaussianhaircut,gaussianhair,groomcap,pan2025hairgs}, for refinement we transform each strand in $\Gamma$ into a chain of cylindrical Gaussians, where each segment is represented by a single Gaussian.
%While earlier hair capture pipelines~\cite{sklyarova2023neuralhaircut,takimoto2024drhair} often relied on soft rasterization~\cite{liu2019soft}, recent Gaussian-based hair renderers~\cite{gaussianhaircut,gaussianhair,groomcap,pan2025hairgs} suggest that strand-aligned Gaussian chains provide an effective differentiable representation for explicit hair geometry. 
This representation is a particularly good fitting in our context, since facial hair often requires fine-grained local corrections---including segment-level geometric updates which will be called in our paper as \textit{local surgery}---which are naturally supported when each strand is represented as an ordered chain of local Gaussian primitives. 

%Given a traced strand polyline $\{\mathbf{v}_0,\ldots,\mathbf{v}_N\}$, where $\mathbf{v}_0$ denotes the root vertex, we place one Gaussian on each segment $(\mathbf{v}_i,\mathbf{v}_{i+1})$, with center $\boldsymbol{\mu}_i=\tfrac12(\mathbf{v}_i+\mathbf{v}_{i+1})$, tangent direction (maximum principal axis) $\mathbf{t}_i=\mathbf{v}_{i+1}-\mathbf{v}_i$, and anisotropic scales $\boldsymbol{\sigma}_i=(\|\mathbf{t}_i\|,\epsilon,\epsilon)$, where $\epsilon$~(=10\,$\mu m$) is a fixed transverse radius. Since the strand vertices are uniformly sampled after tracing, each Gaussian has equal principal-axis length. This yields a strand-aligned differentiable representation where each Gaussian directly corresponds to a strand segment, enabling segment-level gradient updates for geometric refinement.
% During refinement, we intentionally restrict appearance to SH degree $0$. This prevents photometric ``shortcuts'' in which view-dependent shading absorbs geometric errors, forcing the optimizer to explain mismatches primarily through geometry and density rather than specular or other view-dependent effects.

\paragraph{Subject-wise VAE shape prior.}
Instead of optimizing vertex positions \citep{rosu2022neuralstrands,gaussianhaircut}, which is difficult to stabilize in sparse regimes such as facial hair, we opt for building a prior for refinement. However, unlike scalp hair, there is no large-scale dataset to train a generic VAE prior. %\citep{hu2015uschair}. 
%Optimizing free per-vertex directions (e.g., a directional ODE-style representation as in~\citet{rosu2022neuralstrands}, which~\citet{gaussianhaircut}  follows) is brittle for facial hair: it typically requires additional curvature/direction regularization and is difficult to stabilize in sparse regimes. 
%Moreover, unlike scalp hair, there is no large-scale facial hair dataset to train a generic VAE prior, as exists for scalp hair~\cite{hu2015uschair}. 
Following~\citep{ct2hair, groomcap}, we instead train a \emph{subject-specific} 
VAE $E(\cdot)$ from grounded strands and parameterize strand shape through its latent code. We use the VAE architecture from~\citet{rosu2025difflocks}. 
This provides a compact, stable shape space for optimization, and acts as denoiser removing jitter and small tracing artifacts. 

\paragraph{Strand surgery.}
Because facial hair is highly sparse and spatially heterogeneous, a traced strand can pass through dense and empty regions, and even portions on the root-side may need to remain sparse. Consequently, a single root--tail cut-point as the one used in previous work on scalp hair reconstruction \citep{groomcap} is too inflexible for our setting. We instead parameterize each strand with $K$ (where $K < N$) opacity control points, linearly interpolate opacity along the polyline, and prune all segments whose interpolated opacity falls below a threshold. This enables fine-grained local removal and can split one strand into multiple surviving components.

%\paragraph{Multiple opacity parameterization.}
%Among existing approaches, GroomCap~\cite{groomcap} is the only method that explicitly controls strand density during refinement, by learning a minimum transverse radius $\epsilon$ and triggering splitting when scale or opacity exceeds a threshold. It further uses only two opacity values per strand---$o_1$ for the root-side segments and $o_2$ for the final $N_t{=}8$ tail segments---which effectively imposes a coarse root--tail transparency prior that tapers opacity toward the tip. Such a design is suitable for scalp hair, where visually plausible tip attenuation is important and the main goal is to preserve natural strand appearance under volumetric refinement.
%For facial hair, however, our objective extends beyond appearance refinement: we also perform mask-aware \emph{strand surgery}. Because facial hair is highly sparse and spatially heterogeneous, a traced strand may pass through both supported and unsupported regions, and even root-side portions may need to remain sparse. Consequently, a single root--tail cut-point is too inflexible for our setting. We instead parameterize each strand with $K$ (where $K < N$) opacity control points, linearly interpolate opacity along the polyline, and prune all segments whose interpolated opacity falls below a threshold. This enables fine-grained local removal and can split one strand into multiple surviving components.

%\paragraph{Strand surgery.}

For each strand polyline $\mathbf{\gamma_i}=\{\mathbf{p}_0,\ldots,\mathbf{p}_{N-1}\}$, we associate $K$ opacity control points $\mathbf{o}=\{o_k\}_{k=0}^{K-1}$ placed at equally spaced arc-length positions along the strand. For each segment $\mathbf{s}_i=(\mathbf{p}_i,\mathbf{p}_{i=1})$, we compute an opacity $\alpha_i$ by linearly interpolating the neighbor control points at the segment midpoint, and remove segments with $\alpha_i < \tau_\alpha$, with $\tau_\alpha = 0.1$ an empirically defined threshold.

\setlength{\intextsep}{3pt}
\setlength{\columnsep}{9pt}
\begin{wrapfigure}[10]{r}{0.22\textwidth}
    \vspace{-1pt}
  \begin{center} 
    \includegraphics[width=0.22\textwidth]{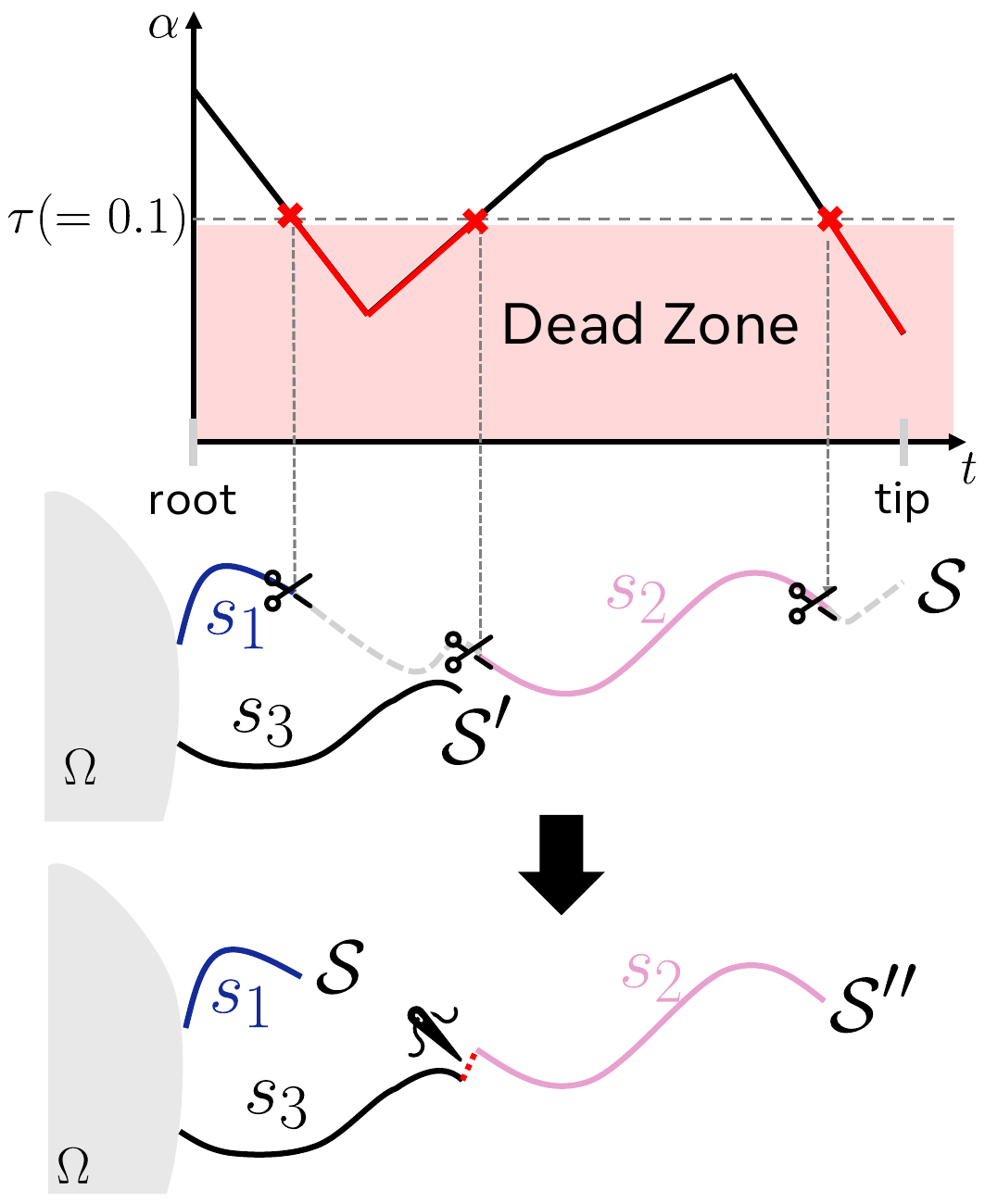}
  \end{center}
  \vspace{-1pt}
  % \caption{\textbf{Hollow illusion}}
  % \label{wrapfig:hollow}
\end{wrapfigure}
This partitions the strand into $M$ surviving discontinuous sub-strands $S=\{s\}_{j=0}^{M-1}$, where $s_j$ is considered root-connected if it contains the root vertex $\mathbf{p}_0$, and floating otherwise. 
For floating surviving sub-strand, we first attempt to re-ground them onto the tracked mesh (\Cref{sec:grounding}). If grounding succeeds, the piece is encoded directly; otherwise, we fall back to stitching (\Cref{sec:tracing}) and reconnect it to a root-connected set before encoding.
Each surviving sub-segment is thus encoded into the latent space of our subject-specific VAE as
\begin{equation}
\hat{\mathbf{z}} = 
\begin{cases}
E(s_i) & m_0, \\
E(\texttt{Ground}(s_i)) & \neg m_0 \wedge \texttt{Ground}(\cdot) \text{ succeds}, \\
E(\texttt{Stitch}(s_i)) & \neg m_0 \wedge \texttt{Ground}(\cdot) \text{ fails}.
\end{cases}
\end{equation}
where $m_0 = (\alpha_0 \ge \tau_\alpha)$ is a boolean mask indicating whether $s_i$ is root-connected, and $\texttt{Ground}(\cdot)$ and $\texttt{Stitch}(\cdot)$ are the grounding and stitching procedures defined in  \Cref{sec:grounding} and \Cref{sec:tracing} respectively. The resulting latent codes are used for subsequent refinement. 

\paragraph{Training.}
We use the same loss as in Stage~I (\Cref{eq:loss_gs}), additionally introducing a latent regularization term $\mathcal{L}_{\mathbf{z}}$ to stabilize refinement under strand surgery, defined as
\begin{equation*}
\mathcal{L}_{\mathbf{z}}=\left\lVert \mathbf{z}-\hat{\mathbf{z}}\right\rVert_1,
\end{equation*}
with $\hat{\mathbf{z}}$ the reference latent obtained after each surgery step, which we compute every 2K iterations, omitting the final iteration to prevent over-pruning, over a total of 10k iterations. This encourages the optimization to remain close to the strand manifold while still allowing photometric refinement.

% \section{Appearance Matching for Facial Hair}
% \carlos{TBD: if it does not provide too much technical insights, we might briefly explain in the results section}

\section{Experiments} 
Here we analyze our method against existing baselines in two different datasets: a) A synthetic high-quality human head dataset, and b) a captured dataset~\cite{beelerfacialhair}. Details of both datasets can be found in \Cref{sec:datasets}. For visualization, we try to minimize visual deception: Comparisons are primarily shown using 3D direction-encoded rendering, and for regions such as beards where local sparsity and growth trends matter more than per-strand quality, we apply a consistent set of arbitrary matte BSDF parameters across all methods. For application results, we render using a production hair BSDF \citep{chiang2015practical} with constant parameters, under a uniform studio-capture lighting setting. The optimization cost and statistics are reported in \Cref{tab:timings_stats}.

%\adrian{Do we have timings?} \jesse{I measure with timestamps! would you refer the table1, below?}

\begin{table*}[t]
  \caption{Per-region statistics and runtimes, computed on a single NVIDIA H200 (144\,GB). 
  \#Gaussians is the output of Stage~I, and \#Strands the final strand count after mean-shift fusion, tracing, stitching, and grounding (S2--S3). Note that Stage IV is skipped for both eyebrow and eyelash regions.}
  \centering
  \small
  \setlength{\tabcolsep}{5pt}
  \begin{tabular}{lcccccc}
    \toprule
    Region & \#Gaussians & \#Strands & Stage~I & Stage II \& III & Stage~IV & Total \\
    \midrule
    Eyelash                  & $\sim$10--100\,K     & 0.3--2.1\,K  & $\sim$15--20\,min   & $<$2\,min     & ---           & $\sim$20\,min   \\
    Eyebrow                  & $\sim$0.15--0.27\,M  & 1.0--4.9\,K  & $\sim$20--30\,min   & $<$3\,min     & ---           & $\sim$30\,min   \\
    Sparse beard (stubble)   & $\sim$0.8\,M         & 4--6\,K      & $\sim$30--50\,min   & $<$5\,min     & $\sim$2\,h    & $\sim$3\,h      \\
    Dense beard (regular)    & $\sim$1--3\,M        & 10--45\,K    & $\sim$80--100\,min  & $\sim$10\,min & $\sim$3\,h    & $\sim$4--5\,h   \\
    Long/full beard          & up to 4.2\,M         & 23--60\,K    & $\sim$100--120\,min & $\sim$10\,min & $\sim$3\,h    & $\sim$5\,h      \\
    \bottomrule
  \end{tabular}

  \label{tab:timings_stats}
\end{table*}

% Note that the 2D directional fields are all extracted from Gabor filter. We will describe the specification of this parameter in the supplementary section.
%\subsection{Baselines}
%We compare against three baselines. For our in-house synthetic dataset, we evaluate against LPMVS~\cite{lpmvs}, a multi-view stereo-based line reconstruction method adapted for hair, and HairGS~\cite{pan2025hairgs}, a 3DGS-based hair reconstruction method conceptually closest to ours. Both are prior-free. 
%
%Additionally, we compare against~\citet{beelerfacialhair} using the inputs and results provided by the authors, as their capture setup is not reproducible in our pipeline. \textbf{Note that only our method includes a grounding stage, which enables downstream applications unavailable to the baselines, including strand implanting, terrain-based pruning, cutting, and facial animation.}
% Beyond geometric applications, the recovered strand asset is compatible with physically based rendering via in-house BSDF parameters, enabling relighting under arbitrary lighting conditions.

\begin{figure*}[h]
    \centering
    \includegraphics[width=\textwidth]{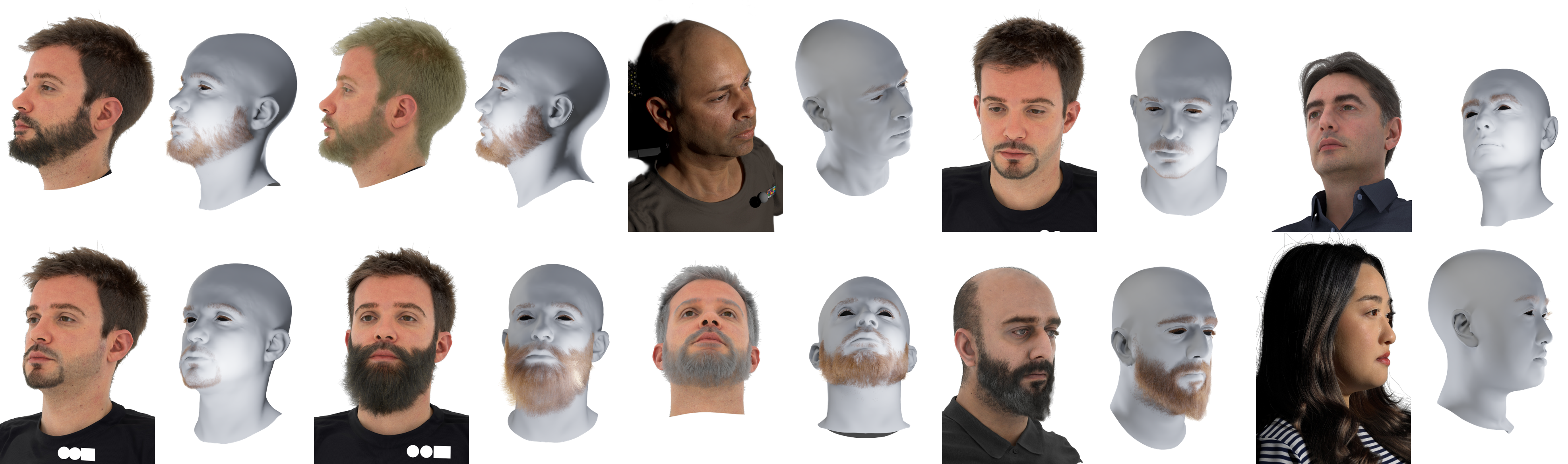}
    
    \caption{\textbf{Reconstruction results.} Our method preserve subject-specific traits including the overall shape and orientation, but also strand sparsity and clustering patterns across diverse facial hair — from fine eyelashes, to short stubble and sparse brows, to long dense beards. \textbf{We recommend readers to zoom-in.}}
    \label{fig:all_subjectViz}
\end{figure*}

\subsection{Qualitative Results}
\Cref{fig:teaser,fig:all_subjectViz} demonstrate our results across a diverse range of facial hair types and subjects. Our method handles the full spectrum of facial hair--from short stubble and medium-length beards to extremely long beards, as well as eyebrows and eyelashes--using a single unified pipeline. Our sparsity-aware formulation naturally accommodates spatially varying density, recovering plausible strand geometry even in sparse regions where prior methods tend to produce diffuse artifacts, and at the same time our method is able to reconstruct the invisible inner parts of dense beards. As shown in the supplemental results, our method also performs robustly on dyed and dark facial hair, demonstrating generalization beyond the limited color range assumed by prior approaches.

\paragraph{Comparisons}

\begin{figure}[h]
    \centering
    \includegraphics[width=0.8\linewidth]{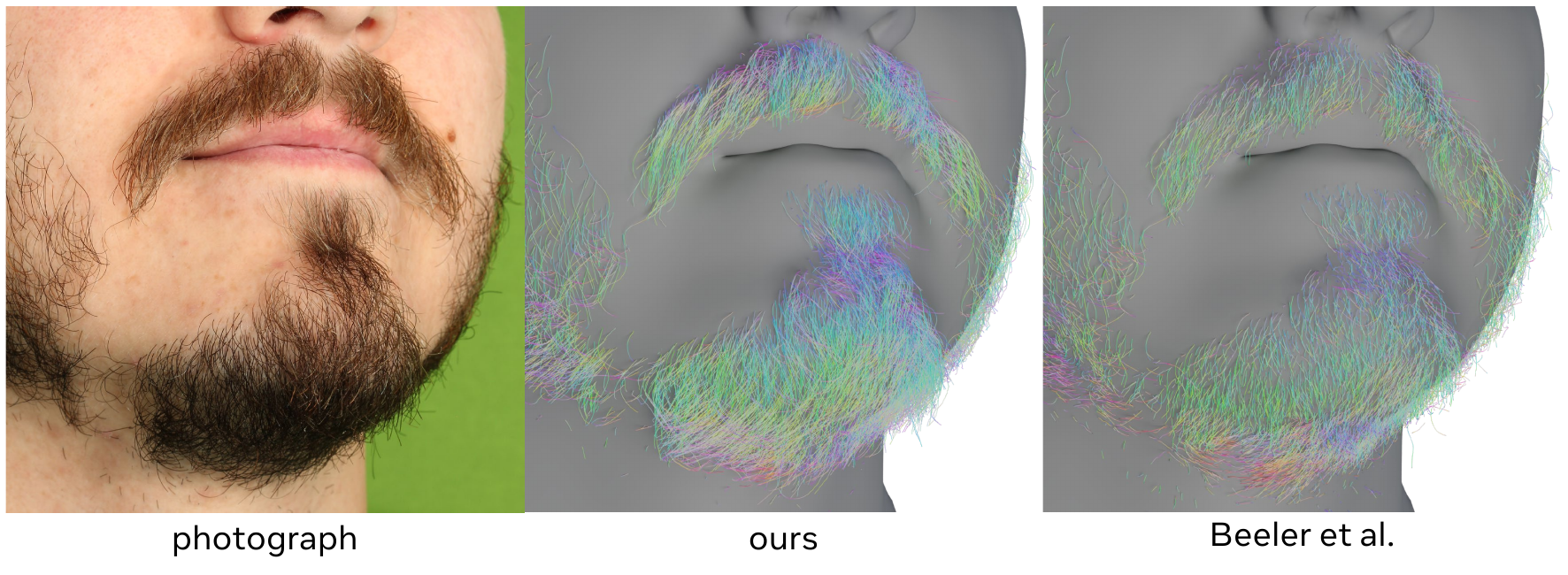}
    \caption{\textbf{Comparison with \citet{beelerfacialhair}.} Using inputs and results provided by the authors.~\citet{beelerfacialhair} recover isolated strands more precisely in sparse regions, while our method achieves more uniform coverage and directional coherence in dense areas such as the chin beard. }
    \label{fig:comparison_beeler}
\end{figure}

\Cref{fig:comparison_beeler} shows a comparison against the method of \citet{beelerfacialhair} in their captured dataset. Our method is able to capture both dense and sparse areas better while providing better directional coherence in dense regions. Note that only our method includes a grounding stage, which enables downstream applications unavailable to the baselines, including strand implanting, terrain-based pruning, cutting, and physics simulation.

\begin{figure}[!h]
    \centering
    \includegraphics[width=0.86\textwidth]{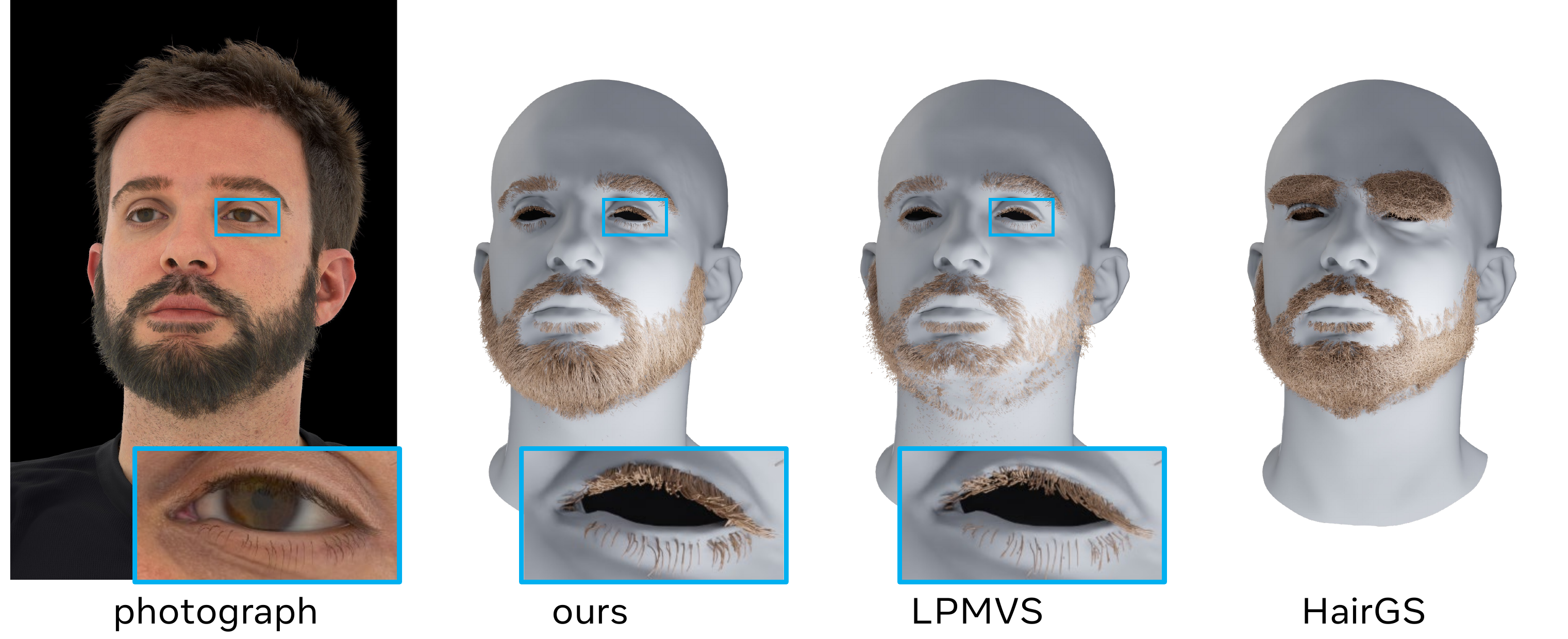}
    \includegraphics[width=0.86\textwidth]{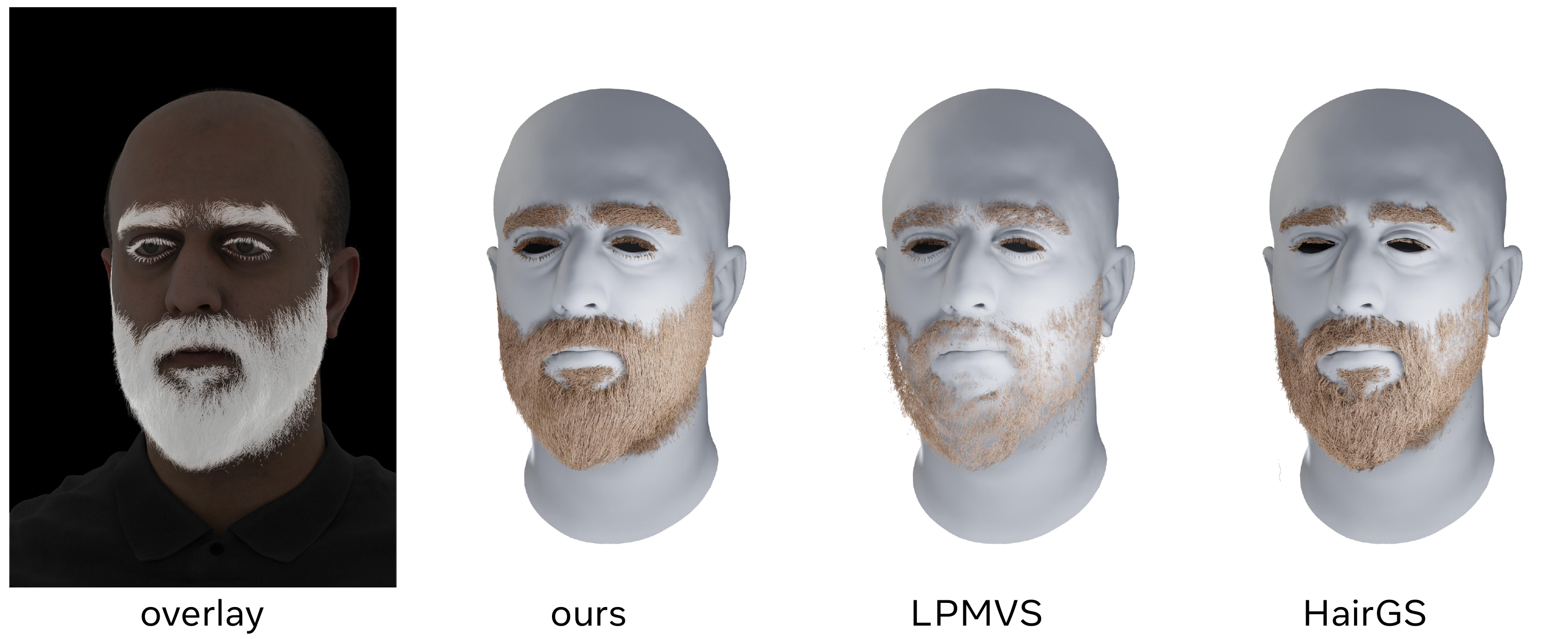}    
    \caption{\textbf{Qualitative comparison on bearded subjects.} Our method recovers coherent strand geometry across both beard and eyebrow regions, faithfully recovering the sparse, spatially varying facial hair distribution, while LPMVS and HairGS produce noisier and less faithful strand coverage. 
    In addition, our method is able to capture fine eyelash strands (blue insets, top) that are missed or poorly reconstructed by LPMVS. HairGS fails to reconstruct the eyebrow region entirely, producing a flat, undifferentiated patch. }
    \label{fig:comp_all_carlos}
\end{figure}

\begin{figure}[!t]
    \centering
    \includegraphics[width=1.0\textwidth]{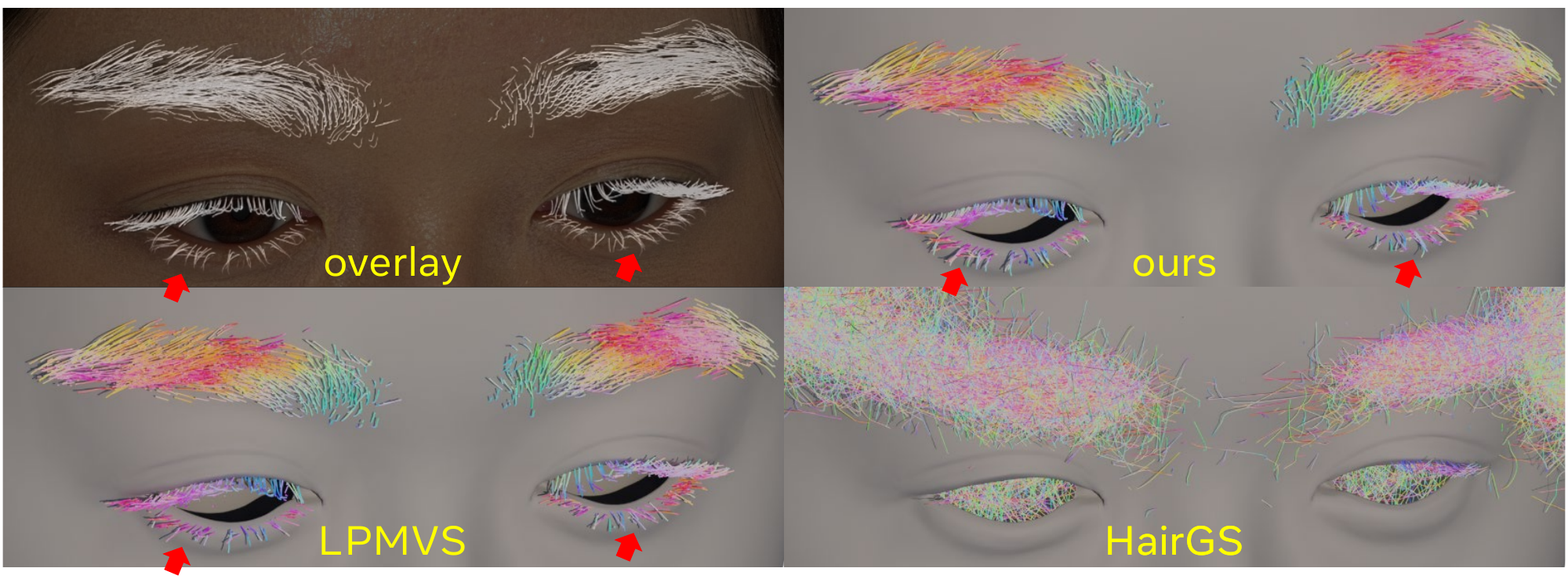}
    \caption{\textbf{Comparison on eyebrows and eyelashes.} Our method outperforms both LPMVS and HairGS on our in-house eyelash and eyebrow dataset, faithfully reconstructing their detailed fine structures.}
    \label{fig:comp_inhouse_eb_el}
\end{figure}

\Cref{fig:comp_all_carlos} show comparisons between our method and two state-of-the-art prior-free methods, HairGS~\cite{pan2025hairgs} and LPMVS~\cite{lpmvs}. Our method is able to accurately represent the spatially-varying distribution of hair, in both sparse and dense areas of the beard, and at the same time produces high-quality reconstruction in lashes and brows. This is further demonstrated in \Cref{fig:comp_inhouse_eb_el}, where we focus the comparison in the eye region. 

%In Fig.~\ref{fig:results} \adrian{Missing?}\jesse{We are waiting for appearance matching results for this figure}, we demonstrate our results across a diverse range of facial hair types and subjects. Being prior-free, our method handles the full spectrum of facial hair — from short stubble and medium-length beards to extremely long beards, as well as eyebrows and eyelashes — within a single unified pipeline. Our sparsity-aware formulation naturally accommodates spatially varying density, recovering plausible strand geometry even in sparse regions where prior methods tend to produce diffuse artifacts. Beyond visible surfaces, the volumetric nature of 3DGS allows our method to recover strand structures in occluded inner regions, such as the underside of beards and eyebrow roots, which are rarely addressed in previous works. Our method also performs robustly on dyed and dark facial hair, demonstrating generalization beyond the limited color range assumed by prior approaches. Across all cases, the reconstructed strands exhibit consistent local orientation, smooth continuity, and physically plausible root attachment, making them directly compatible with downstream editing and animation applications.

% \clearpage
\subsection{Ablation studies}
Here we demonstrate the effect of the techniques implemented in each of the phases of our pipeline. \Cref{fig:overview} shows the aggregated effect of each phase on the reconstruction. 

\begin{figure*}[h]
    \centering
    \begin{subfigure}[t]{0.63\linewidth}
        \centering
        \includegraphics[width=\linewidth]{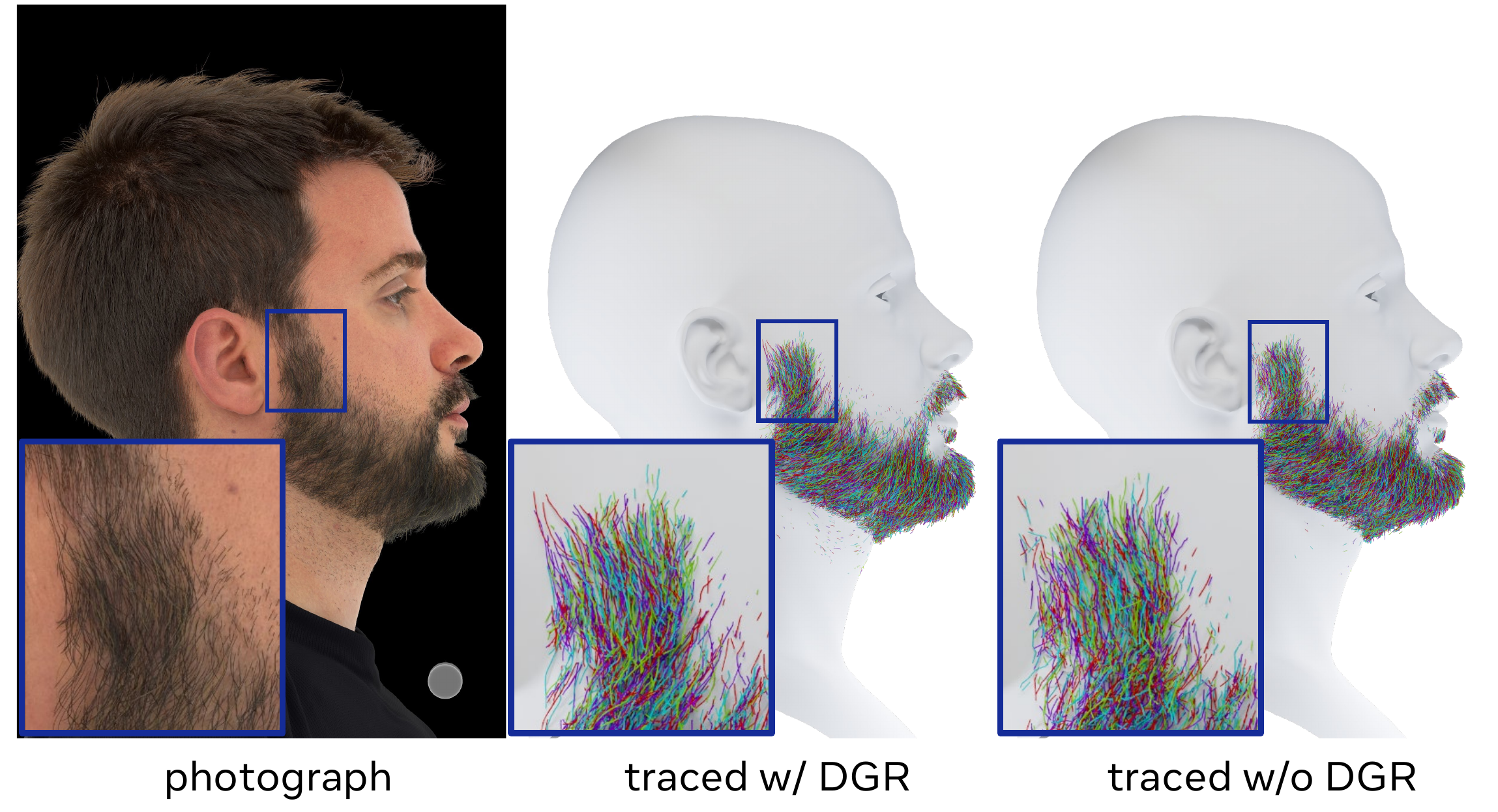}
        \caption{Depth-gated rasterizer (DGR).}
        \label{fig:ablation_dgr}
    \end{subfigure}
    \hfill
    \begin{subfigure}[t]{0.35\linewidth}
        \centering
        \includegraphics[width=\linewidth]{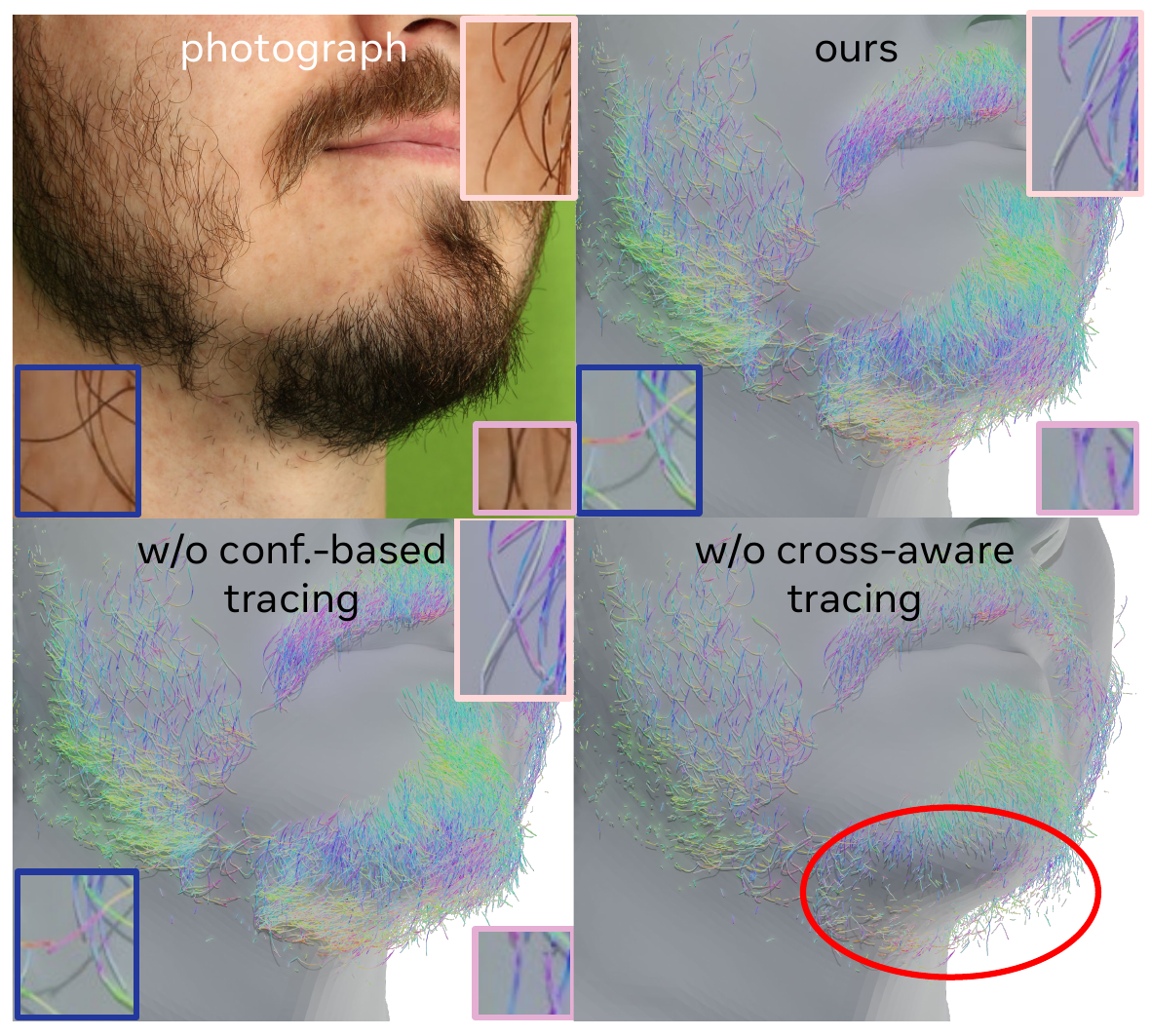}
        \caption{Strand tracing.}
        \label{fig:ablation_tracing}
    \end{subfigure}
    \caption{\textbf{Ablation studies of two pipeline stages.}
    \textbf{(a)} Without DGR, the optimization cannot clearly separate subsurface facial regions from valid strand geometry, causing spurious Gaussians to emerge beneath the surface. These artifacts interfere with strand tracing and connectivity recovery.
    \textbf{(b)} From left to right: input photograph; our traced strands; traced strands without alpha-based mean-shift filtering; and traced strands without cross-aware candidate removal. Without filtering, strands are often misconnected, while without cross-aware removal the recovered strands are sparser with frequent crossings in both beard and mustache regions.}
    \label{fig:ablation_dgr_tracing}
\end{figure*}

\paragraph{Depth-Gated Rasterization (DGR).}
\Cref{fig:onlyhairgaussians} shows that our depth-gated optimization produces a substantially cleaner Gaussian field compared to a standard reconstruction. Without depth gating, the optimizer distributes semi-transparent primitives across both the hair and skin regions, resulting in scattered noise and floating artifacts. In contrast, our gated formulation confines Gaussians strictly to the facial-hair support, yielding primitives more coherently organized along strand trajectories. As shown in \Cref{fig:ablation_dgr}, this cleaner Gaussian field directly translates to improved tracing quality: Spurious sub-surface Gaussians act as false seeds and corrupt the Euler integration, producing chaotic strand directions and connectivity failures, whereas our method recovers coherent, physically-plausible growth patterns throughout the beard region.

\paragraph{Confidence-based and crossing-aware tracing.}
\Cref{fig:ablation_tracing} presents ablation results for our tracing stage, removing two key parts of the method. Without confidence-based mean-shift and tracing, strands are frequently misconnected or disconnected, as low-opacity primitives introduce unreliable directional cues that corrupt Euler integration. Without crossing-aware suppression the redundant parallel candidates are indiscriminately consumed during tracing, resulting in much sparser strand coverage with frequent crossing artifacts, particularly pronounced in the dense beard region. Our full method combines both components, recovering coherent well-connected strands even in dense regions with frequent local crossings.

\begin{figure*}[!h]
    \centering
    \begin{subfigure}[t]{0.49\linewidth}
        \centering
        \includegraphics[width=\linewidth]{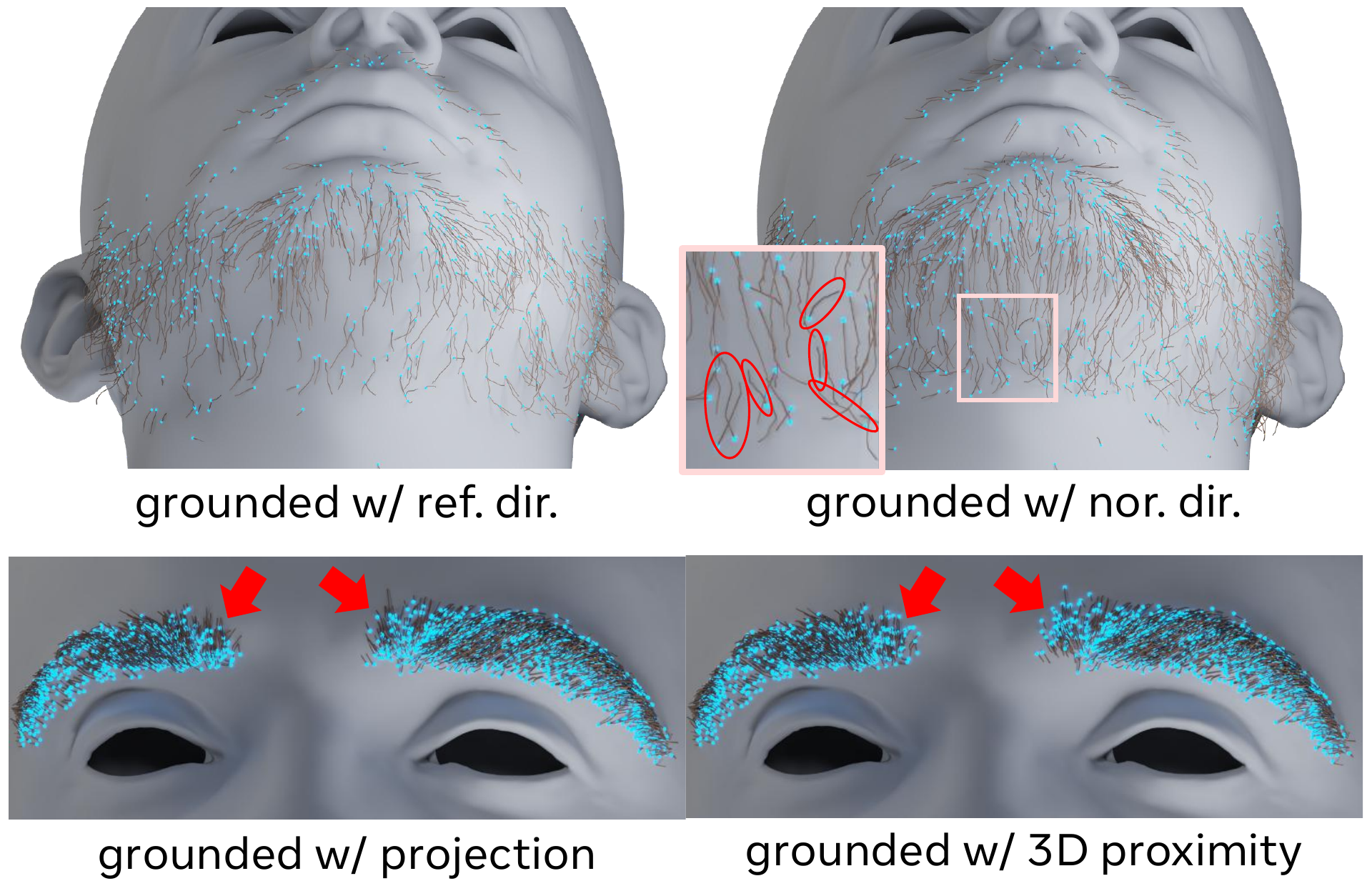}
        \caption{Grounding stage.}
        \label{fig:ablation_grounding}
    \end{subfigure}
    \hfill
    \begin{subfigure}[t]{0.49\linewidth}
        \centering
        \includegraphics[width=\linewidth]{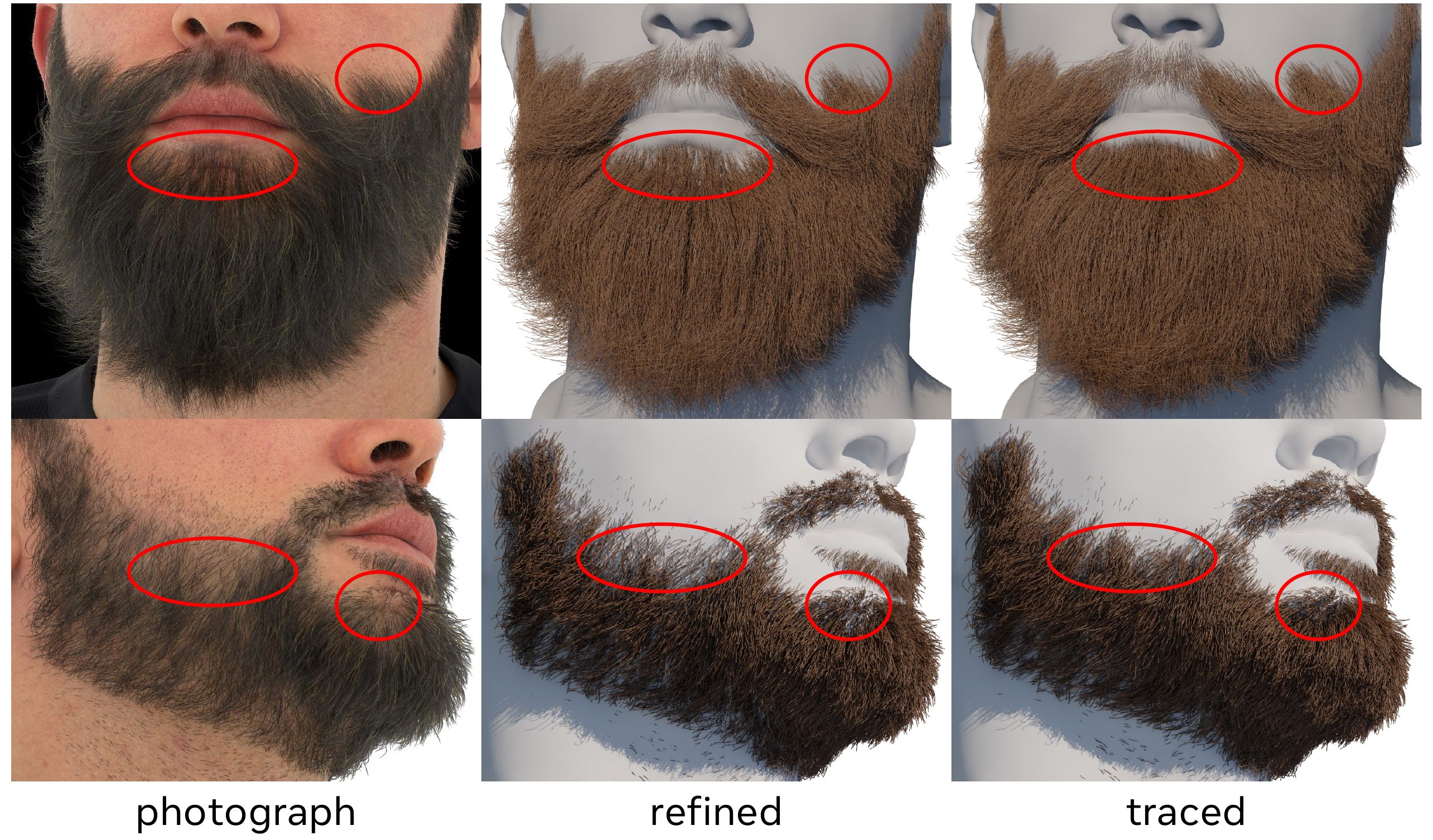}
        \caption{Strand refinement.}
        \label{fig:ablation_refinement}
    \end{subfigure}
    \caption{\textbf{Ablation studies of grounding and refinement stages.}
    \textbf{(a)} \emph{Top:} Our gravity-blended reference direction (left) produces well-aligned strand orientations, whereas relying solely on the surface normal direction~\cite{ct2hair} (right) leads to twisted and misaligned strands, as highlighted by the red ellipses. \emph{Bottom:} Our projection-based grounding (left) yields physically plausible results, whereas naively applying 3D proximity (right) causes strands near the glabella to grow downward, as indicated by the red arrows.
    \textbf{(b)} Our refined result (middle) better captures spatially-varying strand density compared to the raw traced result (right). In sparse regions such as the mustache and lower chin area (red circles), naive tracing tends to over-reconstruct producing unnaturally uniform coverage. In contrast, our refinement stage adaptively prunes redundant strands to match the observed sparsity.}
    \label{fig:ablation_grounding_refinement}
\end{figure*}

\paragraph{Physically-plausible grounding.}
\Cref{fig:ablation_grounding} evaluates our region-adaptive grounding strategy. For beard strands (top row), relying solely on the surface normal direction \citep{ct2hair} introduces twisted and misaligned orientations, as facial hair growth is not perpendicular to the skin surface. Our gravity-blended reference direction resolves this ambiguity, producing well-aligned strand orientations consistent with natural growth patterns. For eyebrow strands (bottom row), naively applying 3D proximity-based grounding \citep{beelerfacialhair} causes strands near the glabella to grow downward (red arrows), which is physically implausible. Our projection-based grounding correctly recovers outward-facing growth directions in this region. These results demonstrate that a single grounding strategy is insufficient across facial hair types.

% \begin{figure}[t]
%     \centering
%     \includegraphics[width=1.0\textwidth]{Figures/Exp/fig_ablation_grounding.pdf}
%     \caption{\textbf{Ablation study of the grounding stage.} 
%     (Top) Our gravity-blended reference direction (left) produces well-aligned 
%     strand orientations, whereas relying solely on the surface normal 
%     direction~\cite{ct2hair} (right) leads to twisted and misaligned strands, 
%     as highlighted by the red ellipses. 
%     (Bottom) Our projection-based grounding (left) yields physically plausible 
%     results, whereas naively applying 3D proximity (right) causes strands near 
%     the glabella to grow in a downward direction, as indicated by the red arrows.}
%     \label{fig:ablation_grounding}
% \end{figure}

 %, and that our region-adaptive formulation is essential for physically plausible results compatible with downstream editing and animation.

\paragraph{Sparsity-aware refinement stage.}
\Cref{fig:ablation_refinement} demonstrates the effect of our sparsity-aware refinement stage. While the tracing stage recovers geometrically-plausible strand trajectories, it produces over-dense regions where redundant, nearly-parallel strands co-occupy the same spatial support. This is particularly problematic for facial hair, where sparsity itself is part of the signal, and sparse patches are often adjacent to dense ones. Our refinement stage allows selectively prunes unsupported segments while preserving well-supported ones. As shown in the red ellipses, the refined result matches better the observed density in sparse regions (e.g., mustache and lower chin), whereas the raw traced result produces unnaturally uniform coverage. The subject-specific VAE further stabilizes this process by projecting strands back onto the learned shape manifold, suppressing the jitter introduced during surgery.

\section{Conclusion}
Facial hair strand-level reconstruction is a long-standing problem, due to its sparse spatially heterogeneous nature, the need of near-strand-level accuracy, and the lack of domain-specific priors, such as those for scalp hair. 
We showed that these challenges can be addressed through a principled four-stage pipeline that takes advantage of 3DGS as a flexible intermediate representation — not only for rendering, but as a substrate for explicit strand extraction, grounding, and density-aware refinement. Our method is prior-free, which allows us to generalize across the full spectrum of facial hair types, from wispy eyelashes to dense beards, while producing grounded strand assets that are directly compatible with downstream editing and animation pipelines. In the~\Cref{appendix:application}, we show additional examples of such applications.

\subsection{Limitations and Future Work}
Our primary limitation stems from two fundamental challenges of photometric optimization. First, \textbf{surface bias}: volumetric rendering concentrates Gaussian density on the outer visible shell~\cite{wang2021neus}, leaving interior beard structures unsupported and causing tracing failures in occluded regions. We discuss these directions in detail in the~\Cref{appendix:more_discussion}.
Second, \textbf{density ambiguity}: controlling strand density through photometric loss alone is fundamentally ill-posed; without knowledge of true strand thickness, volumetric transmittance ambiguity makes it difficult to distinguish over-dense from correctly sparse reconstructions. 
We believe a promising direction is to leverage physically-based differentiable rendering~\cite{jakob2022mitsuba3} under OLAT illumination, where abnormal appearance signals, such as unexpectedly matte regions caused by over-dense strand packing, could directly drive geometry-level pruning. This pipeline would also allow to reconstruct the facial hair appearance, potentially supporting spatially-varying coloration. 
% We discuss these directions in detail in the \textit{supplemental material}.

% \carlos{This method is awesome and great, future work bla bla}

\section{Acknowledgement}

We thank Thrace Kelsick and Wei Wong for their help with synthetic data generation. We are grateful to Olivier Maury, Christophe Hery, Aljaž Božič, Lukas Bode, Matt Jen-Yuan Chiang, and Doug Roble for valuable discussions and feedback. Finally, we thank Ronald Mallet for his leadership and continued support throughout this project.

\section{Ethics Statement}
\label{ethics}

We clarify that, except for those in the~\citet{beelerfacialhair}. dataset, all characters in this paper are fictional. We strongly condemn any misuse of generative artificial intelligence that could harm individuals or disseminate misinformation. While we acknowledge the potential for misuse in human-centered animation generation, we are dedicated to upholding the highest ethical standards in our research. This commitment includes strict adherence to legal frameworks, respect for privacy, and a focus on promoting the generation of positive and constructive content. The use of this data in our work has undergone legal review and received approval from the legal team of our organization.

\section{In-depth Discussion}\label{appendix:more_discussion}

\begin{figure}[!htbp]
    \centering
    \includegraphics[width=0.9\linewidth]{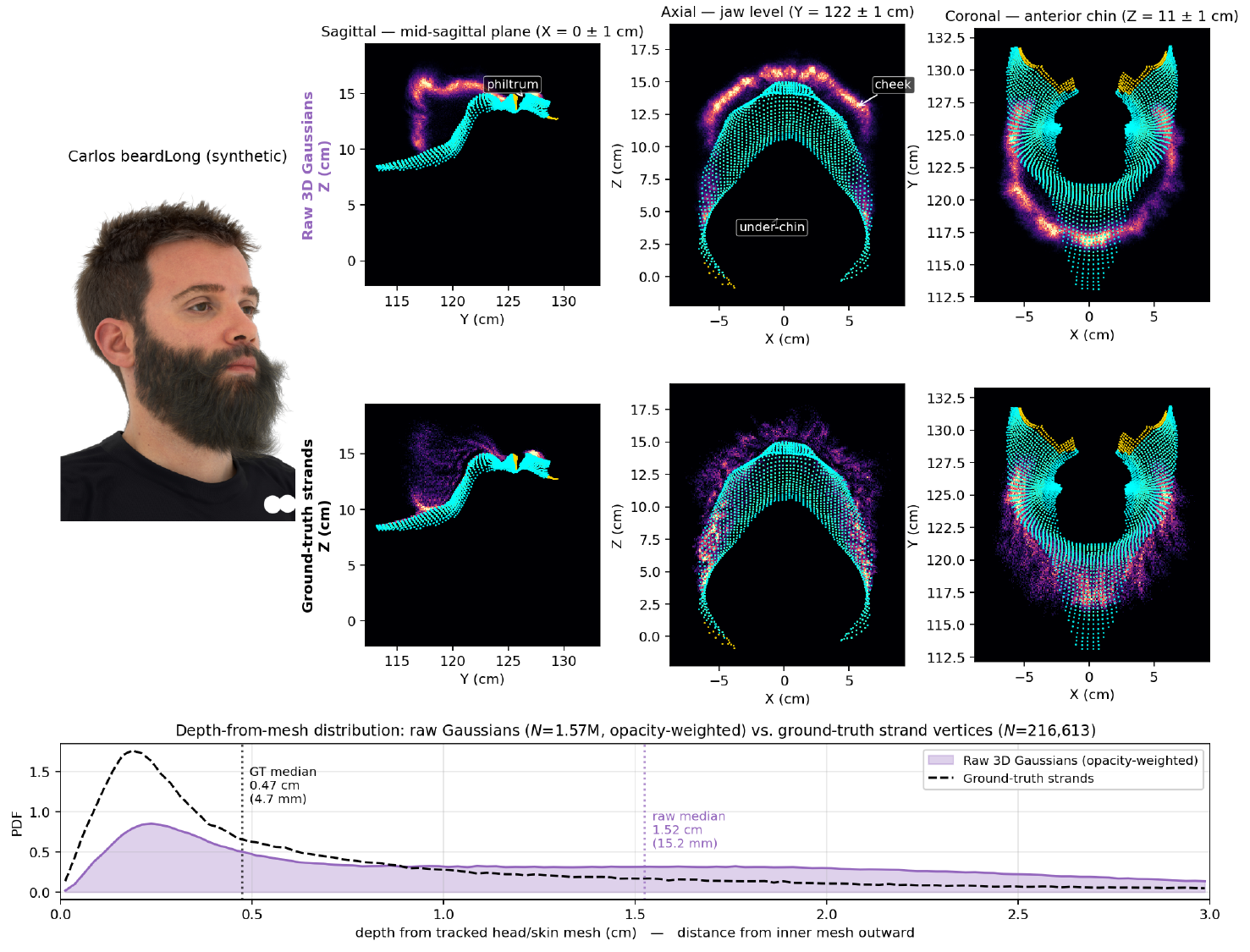}
    \caption{\textbf{Tomographic analysis of raw Gaussian distribution.} Raw Gaussians from photometric optimization concentrate on the outer visible shell, leaving the beard interior unsupported. Tomographic slices of 1,572,297 raw Gaussians along three anatomical planes show that density forms a thin outer band, while the philtrum, cheeks, and underside of the chin remain visibly sparse. This is quantitatively confirmed by the depth distribution: ground-truth strands fill the interior densely (median 4.7 mm from mesh; 35.1\% within 3 mm), whereas optimized Gaussians are pushed outward by roughly $3\times$ (median 15.2 mm; only 12.7\% within 3 mm), producing the tracing failures discussed above.}
    \label{fig:fig_discussion_biasedVolume}
\end{figure}

Despite our sophisticated tracing, grounding, and refinement stages, our method occasionally struggles in regions where the 3D orientation field becomes unreliable. This is an inherent property of view-dependent volumetric optimization~\cite{wang2021neus, yariv2021volume}: since supervision is driven purely by photometric consistency, density concentrates on the outer visible shell while interior regions remain unsupported. Although this is a general property of volumetric representations, it is particularly problematic for dense beard reconstruction, where the interior strand structure is physically present but consistently occluded from all camera views. As evidenced by the Gaussian tomography in Fig.~\ref{fig:fig_discussion_biasedVolume}, the optimized Gaussians are predominantly distributed in the narrow band between the tracked mesh and the outer shell, leading to tracing failures in poorly observed interior regions such as the philtrum, cheeks, and the underside of the chin.

A promising direction to address this limitation would be to replace the volumetric lifting stage with a continuous neural field representation, which could exploit the spectral bias of MLPs to produce smoother and more complete orientation fields in near-surface regions, combined with a backward growing strategy from the mesh surface outward. However, existing backward growing approaches~\cite{groomcap} are designed for scalp hair, where strands originate from a well-defined scalp surface with known root positions. Facial hair, by contrast, emerges from spatially distributed and semantically distinct regions — eyebrows, eyelashes, and beard — whose root positions are not known a priori, making scalp-surface-anchored initialization inapplicable. We leave the design of a facial-hair-aware backward growing scheme as a promising direction for future work.

An alternative direction would be to propagate orientation information from the well-supported outer shell into the interior via anisotropic diffusion, analogous to the 2D orientation diffusion used in~\cite{paris08}, but applied directly to the 3D Gaussian field. High-opacity Gaussians on the outer shell could serve as reliable anchors, and a heat-equation-style iterative scheme could fill in the orientation of interior regions that lack direct photometric supervision. We leave this as another promising direction for future work, too.

\section{Additional details}
\subsection{Stage II: Gaussian Filtering}
\label{sec:stageii_denoising}
% \begin{figure}[t]
%     \centering
%     \includegraphics[width=1.0\columnwidth]{Figures/Exp/fig_opacity_viz.pdf}
%     \caption{\textbf{The opacity visualization} shows that low-opacity Gaussians are often aligned with redundant parallel structures or noisy outliers.}
%     \label{fig:viz_opacity}
% \end{figure}

Facial hair presents additional challenges compared with scalp hair. It is typically much sparser and lower-density, while containing frequent local crossings and intersections. As a result, these ambiguities are more exposed in the optimized Gaussian field. Moreover, due to spatially varying Gaussian anisotropy---especially in the two minor axes orthogonal to the principal direction---a finite strand segment cannot be reliably interpreted as a one-Gaussian-per-segment representation. In practice, the field often contains redundant near-parallel supports, short crossing artifacts, and locally inconsistent thickness patterns. 
We therefore first perform \emph{denoising} to consolidate the Gaussian field into cleaner strand-aligned evidence. 

We use the principal (maximum-covariance) axis of each Gaussian as its local strand direction. This orientation is already used during our depth-gated training and is reused in the subsequent denoising/tracing stages. 
Even with mesh-depth-gated training (which suppresses sub-surface Gaussians), the reconstructed facial-hair field still contains nontrivial noise. Facial hair is susceptible to view-dependent specularity, and 3DGS optimization may introduce view-averaging artifacts; as a result, spurious primitives can remain near true hair regions. We therefore first apply a \textbf{filtering} step: we keep Gaussians with opacity $\alpha$ above a predefined threshold and near-needle anisotropy, measured by effective rank~\cite{erank}, i.e., $\mathrm{erank}$ following~\citet{erankGaussian}. In addition, we prune Gaussians whose center $\mu$ lies inside the tracked mesh to prevent penetration.

\setlength{\intextsep}{3pt}
\setlength{\columnsep}{9pt}
\begin{wrapfigure}[8]{r}{0.5\textwidth}
    \vspace{-1pt}
  \begin{center} 
    \includegraphics[width=0.5\textwidth]{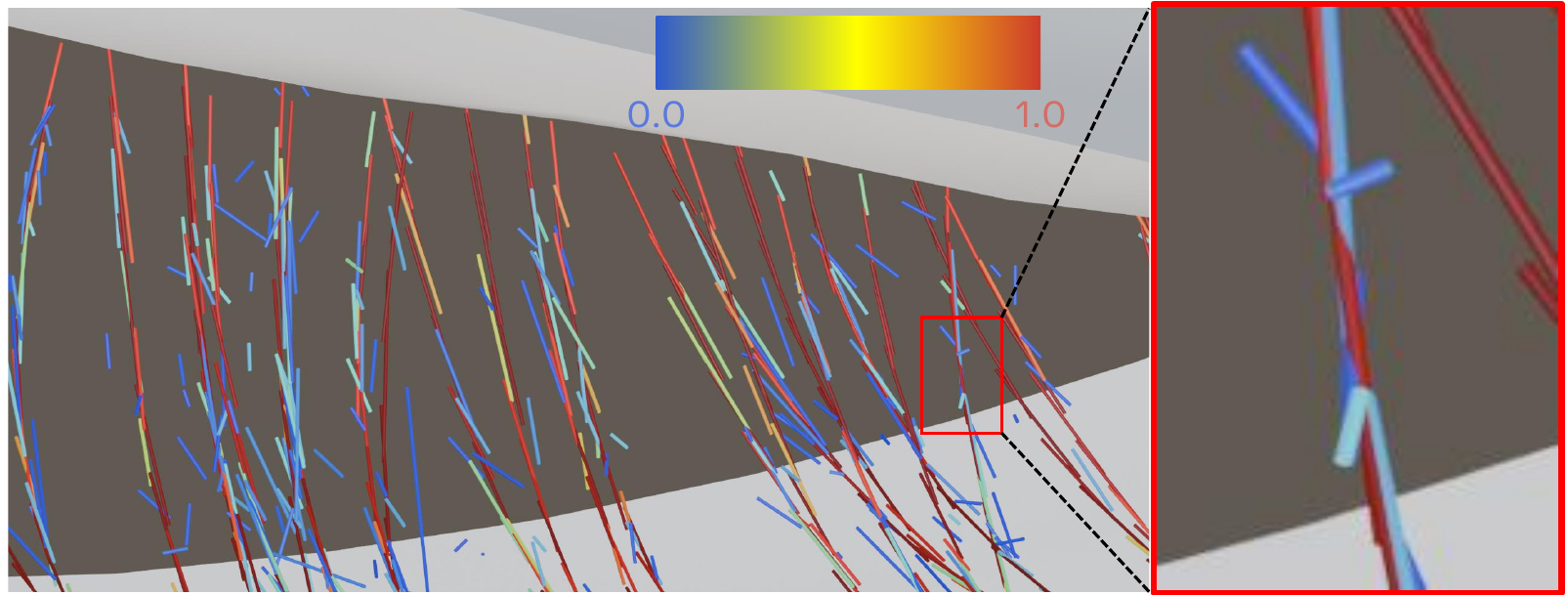}
  \end{center}
  \vspace{-1pt}
  % \caption{\textbf{Hollow illusion}}
  % \label{wrapfig:hollow}
\end{wrapfigure}

When visualizing the filtered Gaussians (inset), we consistently observe that \emph{low-opacity} candidates tend to be less reliable: they often lie slightly off the underlying strand centerline and exhibit larger orientation deviations compared to high-opacity ones. This motivates using opacity as an explicit confidence signal during mean-shift consolidation~\cite{lpmvs}, preventing unstable, low-confidence candidates from biasing the recovered line-like modes.

\subsection{Stage II: Tracing}
\label{sec:tracing_app}
Let each Gaussian candidate be represented as an oriented point $G_i=\{\mu_i, v_i, \alpha_i\}$, where $\mu_i\in\mathbb{R}^3$ is the center, $v_i\in\mathbb{S}^2$ is the maximum principal axis, and $\alpha_i\in[0,1]$ is opacity. Given the current state $P^t=\{p^t,d^t\}$, we form a plane $\Pi_{P^t}$ with normal $d^t$ and treat neighboring candidates as 3D lines to obtain intersection points $x_i=\{\mu_i, v_i\}$ with $\Pi_{P^t}$, searched within radius $r_{\text{nei}}$.
We then update:
\begin{equation}
p^{t+1}=\frac{\sum_{i=0}^{M}\tilde w_i\,\mu_i}{\sum_{i=0}^{M}\tilde w_i},
\qquad
d^{t+1}=\mathrm{normalize}\!\left(\sum_{i=0}^{M}\tilde w_i\,v_i\right),
\label{eq:ms_update_opacity}
\end{equation}
where $i=0$ denotes the self-term ($x_0=p^t$, $v_0=d^t$).
The opacity-weighted bilateral kernel is
\begin{equation}
\tilde w_i
=\alpha_i^{\lambda}
\exp\!\left(
-\frac{\|p^t-x_i\|_2^2}{2\sigma_p^2}
-\frac{\theta_i^2}{2\sigma_d^2}
\right),
\qquad
\theta_i=\arccos\!\left(d^t\cdot v_i\right),
\label{eq:ms_weight_opacity}
\end{equation}
where $\sigma_p$ and $\sigma_d$ are the spatial and angular bandwidth parameters, respectively, and $\lambda$ is the opacity confidence exponent that prevents low-opacity seeds from drifting toward spurious neighbors during mean-shifting. Since the self-term is included, low-opacity seeds have reduced self-support and are naturally pulled toward nearby high-opacity modes, while high-opacity seeds remain stable and are less affected by spurious low-opacity neighbors. This yields a cleaner, less redundant set of consolidated strand directions for downstream tracing.

Let the clean Gaussian set be $\tilde{\mathcal{G}}=\{(\tilde\mu_i,\tilde v_i,\alpha_i)\}$, with $\tilde\mu_i$ its mean and $\tilde v_i$ its principal axis direction. We manage the set of seeds as an opacity-based priority queue, where each Gaussian center serves as a candidate seed and high-opacity primitives are processed first. Any Gaussian consumed either as a seed or as a neighbor during consensus updates is removed from the queue and skipped in future iterations, preventing duplicate traces. Starting from a seed $\mathbf{p}_0$ popped from this queue, we trace bidirectionally — advancing in both $+\mathbf{v}_0$ and $-\mathbf{v}_0$ directions — and concatenate the two resulting trajectories into a single strand. At each step we advance with a forward Euler step
\begin{equation}
\tilde{\mathbf{p}}_{n+1}=\mathbf{p}_n+s\cdot\mathbf{v}_n,
\label{eq:euler_step}
\end{equation}
where $s$ is the step size set to $0.3\,\mathrm{mm}$ for beard and eyebrow regions and $0.1\,\mathrm{mm}$ for eyelashes. Since Gaussian primitives have heterogeneous principal-axis lengths, we uniformly resample each needle along its principal axis into fixed-length segments of $0.1\,\mathrm{mm}$, copying the orientation and opacity attributes to each resampled point. This produces a uniform directional point set that prevents tracing disconnects caused by large inter-Gaussian gaps.

% Starting from a seed position $\mathbf{p}_0$ \adrian{How is this seed position defined and where? Is it at the scalp or where?}\jesse{As I said at the next paragraph, those are popped by opacity-based priority queue!}, we advance with a forward Euler step
% \begin{equation}
% \tilde{\mathbf{p}}_{n+1}=\mathbf{p}_n+s\cdot\mathbf{v}_n,
% \label{eq:euler_step}
% \end{equation}
% where $s$ is the step size set to $0.3\,\mathrm{mm}$ for beard and eyebrow regions and $0.1\,\mathrm{mm}$ for eyelashes.
%
Note that we manage the set of seeds as an opacity-based priority queue and split each Gaussian into unit-length segments of $0.01\,\mathrm{mm}$ to handle potential tracing disconnects. %\adrian{What does this mean? That for each Gaussian we are generating N small Gaussians each 0.01 mm, so we can later do the NN search? } \jesse{I meant, the Gaussian needles have heterogeneous length (principal axis), so I just splited the Gaussians needles to the principal axis with 0.1mm (if needle has 0.5mm -> 5 direction points! like this)} 

We then gather neighboring Gaussian indices
\begin{equation}
\mathcal{N}_{n+1}=
\left\{
k:\|\tilde\mu_k-\tilde{\mathbf{p}}_{n+1}\|\le \tau_r,\;
\angle(\tilde v_k,\mathbf{v}_n)\le \tau_a
\right\},
\label{eq:neighbor_set}
\end{equation}
where $k$ is a Gaussian index in $\tilde{\mathcal{G}}$.
Conventional forward-Euler tracing in prior work~\citep{lpmvs} typically operates on a noisy directional point cloud with limited confidence cues, and thus offers little principled mechanism to down-weight unreliable observations. In contrast, Gaussians provide an additional supervisory signal: per-primitive opacity.  Tracing terminates when no valid neighbors remain, i.e., $\mathcal{N}_{n+1}=\emptyset$.

Empirically, we found that low-opacity Gaussians around a strand often correspond to (i) thin, nearly-parallel primitives that merely support the local mode, or (ii) unreliable directions caused by ill-conditioned anisotropy (e.g., a weak principal axis that yields twisted or unstable orientations). A naive hard threshold on opacity during denoising is undesirable, as it can remove legitimate but weak evidence and break strand connectivity. Instead, we incorporate opacity directly into tracing: at each Euler step, we predict the next position and direction using \emph{opacity-weighted} aggregation of nearby candidates. This soft weighting suppresses noisy contributions while preserving faint but consistent support, leading to more stable trajectories and improved strand continuity. Concretely, we compute the consensus update as
\begin{equation}
\mathbf{p}_{n+1}=
\frac{\sum_{k\in\mathcal{N}_{n+1}}\alpha_k\,\tilde\mu_k}
{\sum_{k\in\mathcal{N}_{n+1}}\alpha_k},
\qquad
\mathbf{v}_{n+1}=
\mathrm{normalize}\!\left(
\sum_{k\in\mathcal{N}_{n+1}}\alpha_k\,\tilde v_k
\right).
\label{eq:alpha_weighted_update}
\end{equation}

% Unlike prior mesh-aware tracing scheme~\cite{groomcap} that steers future growth directions after penetration occurs, our correction acts directly on the updated strand position. This is particularly well aligned with our formulation, since the next point is explicitly estimated from local Gaussian support rather than obtained purely by integrating a direction field. In practice, this position-level correction provides a more direct safeguard against persistent surface penetration while preserving the data-anchored nature of our tracing update.  \adrian{When does the Euler integration stops?} \jesse{I wrote at the previous paragraph, would u check it?TL;DR, if there is no more seeds, stop it :)}

Additionally, to avoid an overlooked point from prior pipelines~\citep{lpmvs,ct2hair,hairinverserendering} that indiscriminately remove all nearby points after tracing one strand segment, we newly suggest a \textbf{crossing-aware suppression} rule, motivated by frequent near-crossings found in facial hair. After accepting segment $\mathbf{s}_n=(\mathbf{p}_n,\mathbf{p}_{n+1})$, we define a local cylinder $\mathcal{C}(\mathbf{s}_n,r_c)$.
Candidates inside the cylinder with small angular deviation are removed as redundant support 
% \adrian{what is $k$? is it $\mu_k$ the mean of the Gaussian?}: \jesse{yes, but as I said before, the gaussian needles are splitted by 0.1mm... so kind of misleading now...}
\begin{equation}
\textcolor{burgundy}{\mathbf{remove}}(k)\iff
k\in\mathcal{C}(\mathbf{s}_n,r_c)\;\wedge\;
\angle(\mathbf{t}_k,\mathbf{v}_n)<\tau_{\mathrm{sim}}.
\label{eq:remove_rule}
\end{equation}
For candidates with larger angular deviation, we perform a short look-ahead trace and preserve them if they yield a valid continuation:
\begin{equation}
\textcolor{darkgreen}{\mathbf{keep}}(k)\iff
k\in\mathcal{C}(\mathbf{s}_n,r_c)\;\wedge\;
\angle(\mathbf{t}_k,\mathbf{v}_n)\ge\tau_{\mathrm{sim}} \wedge\;
\texttt{Trace}_{\mathrm{pseudo}}(k)\ge L_{\min}.
\label{eq:keep_rule}
\end{equation}
This mechanism preserves plausible crossing strands while effectively suppressing duplicated parallel traces, where $\tau_{\mathrm{sim}}$ is the angular similarity threshold distinguishing parallel from crossing candidates, and $L_{\min}$ is the minimum valid strand length for the pseudo trace..
% Finally, we manage seeds with a global max-priority queue keyed by opacity:
% \begin{equation}
% k^\star=\arg\max_{k\in\mathcal{P}_{\mathrm{alive}}}\alpha_k.
% \label{eq:priority_seed}
% \end{equation}
% Prioritizing high-$\alpha$ seeds establishes reliable structures early, improving robustness in sparse facial-hair regions and reducing kinking arising from premature or inconsistent local decisions. The complete procedure is detailed in Algorithm~\ref{alg:strand_tracing}, where 
The \texttt{Trace} function implements the integration scheme described above.
The final output of this tracing is a set of traced strands
$\Gamma=\{\gamma_1,\gamma_2,\ldots,\gamma_{N_s}\}$ with $N_s$ strands,
where each strand $\gamma_i\in\mathbb{R}^{N_k\times 3}$ denotes a 3D polyline with $N_k$ points.

\section{Datasets}
\label{sec:datasets}
\paragraph{High-quality synthetic human head dataset}
We render a high-quality human head dataset authored by professional technical artists, which provides full control over geometry, materials, and visibility cues. All images are rendered in a virtual Blender environment using 20 synchronized cameras at 8K resolution, enabling dense multi-view observations. We used $\times2$ down-sampled images. Since the entire capture process is performed in simulation, camera intrinsics/extrinsics are exact and directly available, and thus no additional calibration procedure is required. Moreover, the dataset provides ground-truth head geometry for each identity. Therefore, unlike real capture pipelines, our setup does not require external head-tracking or geometric alignment steps—both the camera parameters and the head mesh are natively consistent by construction. The dataset consists of five subjects: one female and four males. All subjects have their own eyebrows and eyelashes. One male subject is captured across four hairstyle variations (manbun, undercut, beard, and messy comma), another male subject features a beard only, and the remaining two male subjects have only eyebrows and eyelashes.

\paragraph{Captured dataset}
We preprocess the provided dataset by \citet{beelerfacialhair}, where there is a single subject with thirteen forward facing calibrated cameras with a scanned head mesh. For the upcoming grounding stage, we optimized an in-house 3DMFM supervised by facial keypoints extracted from the images and a scanned mesh. 
The matting mask is extracted by a reimplementation of Beeler´s method~\citet{beelerfacialhair}, with a Gabor filter and the non-maximum suppression algorithm.

\section{Applications}\label{appendix:application}
\subsection{UV terrain-based trimming} 
\setlength{\intextsep}{3pt}
\setlength{\columnsep}{9pt}

\begin{wrapfigure}[10]{r}{0.32\textwidth}
    \vspace{-3pt}
  \begin{center} 
    \includegraphics[width=0.32\textwidth]{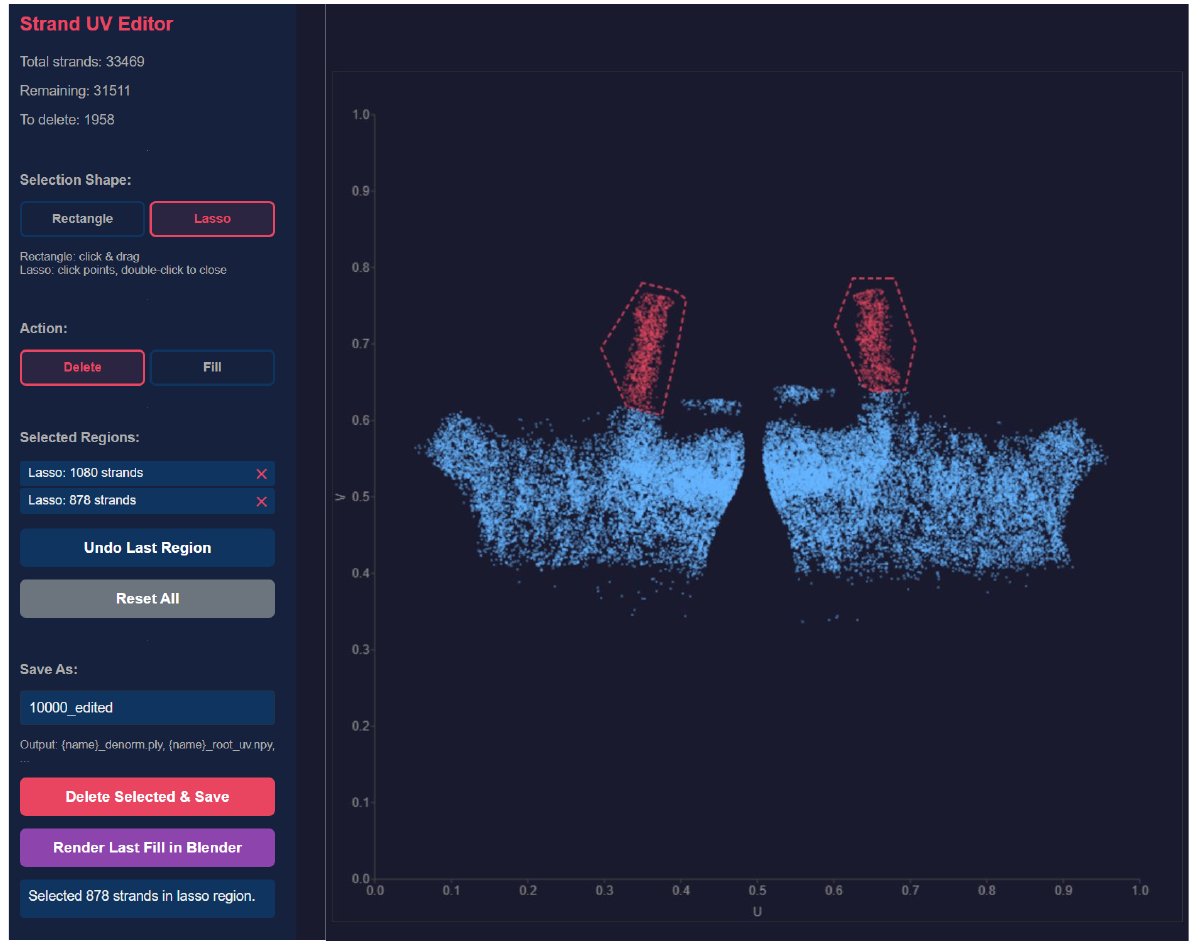}
  \end{center}
  % \vspace{-1pt}
  % \caption{\textbf{Hollow illusion}}
  % \label{wrapfig:hollow}
\end{wrapfigure}
The UV parameterization of the underlying face mesh provides a natural 2D domain for stylistic edits, as every reconstructed strand is indexed by the texel-space coordinate of its root point. We implement an interactive editor that visualizes the root distribution as a 2D density map---a ``UV terrain''---and lets the user select arbitrary strand subsets via rectangle or lasso tools, followed by delete or fill operations (inset). Editing in UV space is markedly more ergonomic than direct 3D manipulation: self-occlusion and viewpoint dependence are eliminated, and spatially coherent regions of the groom map to compact 2D clusters that are easy to isolate. 

\begin{figure*}[h]
    \centering
    \vspace{3mm}
    \includegraphics[width=0.8\textwidth]{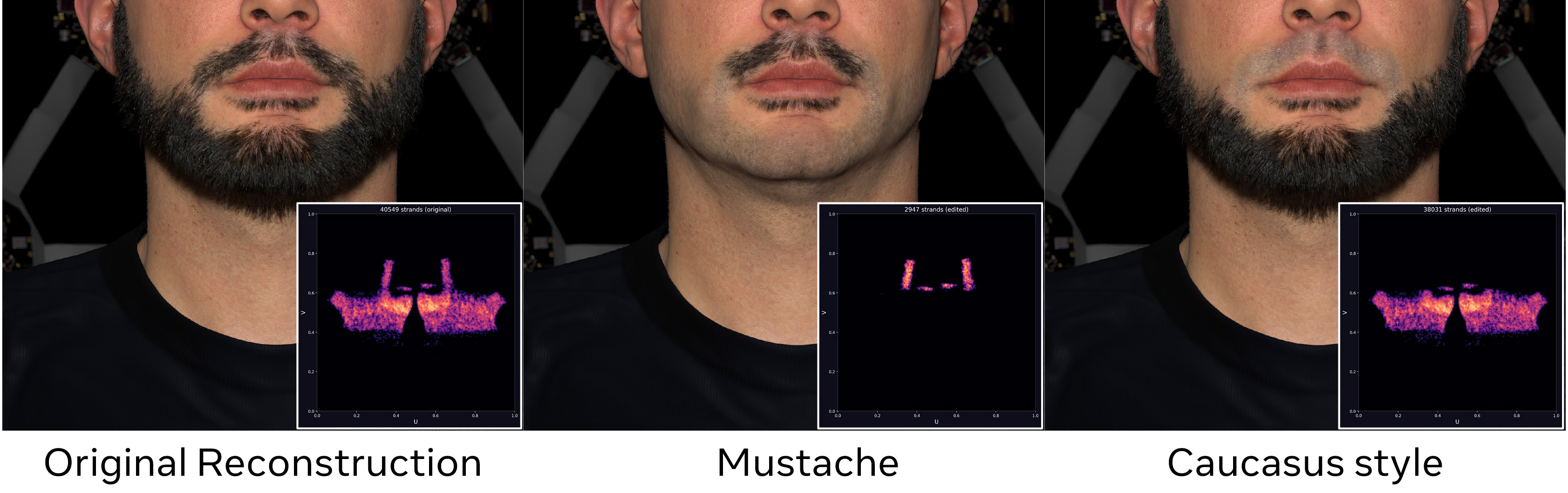}
    \caption{\textbf{UV terrain-based trimming application.} We show the reconstructed geometry (left), the result of UV terrain-based trimming with a mustache (center), and a beard-only style inspired by a Caucasian grooming style (right). Each right-bottom inset is heatmap of root points in texel space.}
    \label{fig:application_uvRemoval}
\end{figure*}

\Cref{fig:application_uvRemoval} demonstrates two stylistic edits obtained from the same reconstruction: a mustache-only variant, produced by deleting the chin and cheek clusters, and a Caucasus-inspired style, in which the mustache region is removed while the cheek and chin growth is preserved. The right-bottom inset of each panel shows the resulting root density in UV space, highlighting which clusters were retained.

% \begin{figure}[!htbp]
%     \centering
%     \includegraphics[width=1.0\columnwidth]{figures/figure_application_uvRemoval_interactive.pdf}
%     \caption{\textbf{UV terrain-based trimming application.} We show the reconstructed geometry (left), the result of UV terrain-based trimming with a mustache (center), and a beard-only style inspired by a Caucasian grooming style (right). Each right-bottom inset is heatmap of root points in texel space.}
%     \label{fig:application_uvRemoval_tool}
% \end{figure}

\subsection{Beard Implantation}
\begin{figure}[!h]
    \centering
    \vspace{3mm}
    \includegraphics[width=0.8\linewidth]{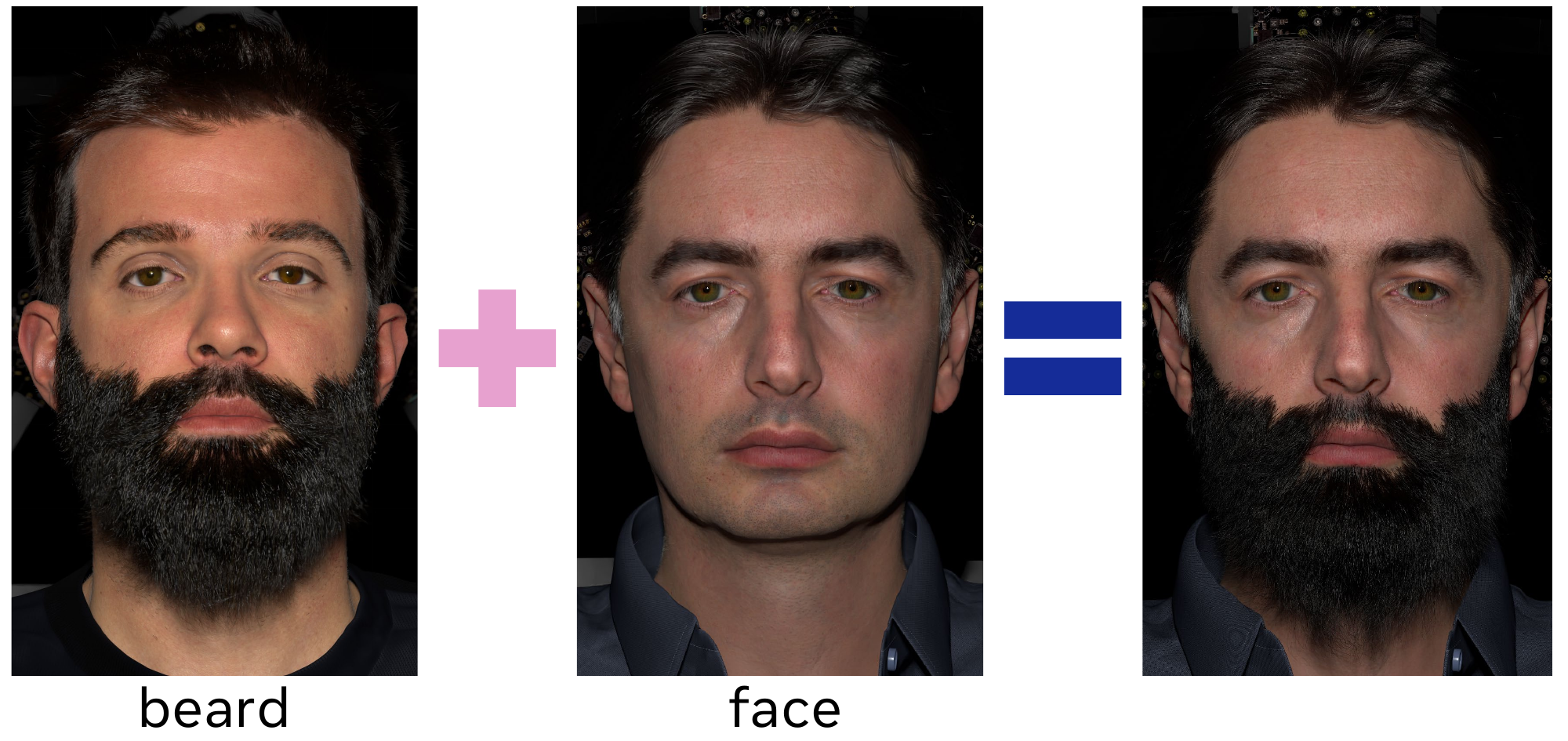}
    \caption{\textbf{Beard implantation.} Because our strands are grounded onto a shared UV topology, the reconstructed facial hair asset can be directly transferred across subjects without re-grounding. Given a source beard (left) and a target face mesh (middle), we implant the strands by mapping root positions through the shared UV space, producing a plausible result that naturally conforms to the target's facial geometry (right).}
    \label{fig:application_implant}
\end{figure}

The shared UV topology of our face mesh enables a particularly simple form of cross-subject transfer: because every reconstructed strand is parameterized by a texel-space root coordinate rather than a subject-specific 3D position, a facial hair asset reconstructed on one subject can be implanted onto another by reusing the same UV root coordinates and re-grounding the strands onto the target geometry. No re-fitting, retraining, or strand-level correspondence is required. ~\Cref{fig:application_implant} demonstrates this on a beard-to-face transfer: the dense beard reconstructed from the source subject (left) is mapped onto a clean-shaven target (middle) through the shared UV space, yielding a result (right) in which the implanted strands naturally conform to the target's facial geometry while retaining the density, length distribution, and growth direction of the source.

\subsection{Eyelash Curling}
As an editing application, we amplify the curvature of each reconstructed eyelash while preserving its attachment to the lid. We approximate every strand $\{\mathbf{p}_0, \ldots, \mathbf{p}_{N-1}\}$ (with $\mathbf{p}_0$ the root) as a planar circular arc: PCA on the centered points yields the best-fit plane, and an algebraic least-squares fit on the 2D projections recovers the arc center $\mathbf{c}$ and radius $R$, from which signed relative angles $\Delta\theta_i$ are extracted. Given a curling factor $\alpha > 1$, we scale the curvature $\kappa = 1/R$ by $\alpha$ through the joint update $R \mapsto R/\alpha$ and $\Delta\theta_i \mapsto \alpha\,\Delta\theta_i$, which leaves the arc length $s_i = R\,\Delta\theta_i$ of every segment invariant; by placing the new center along the original root-to-center direction, the root position, the root tangent, and the curvature sign are all preserved exactly. The resampled 2D points are lifted back to 3D through the PCA basis, followed by a rigid translation that snaps the root onto $\mathbf{p}_0$ to absorb residual planarity error. As ~\Cref{fig:application_eyelashCurling} shows, this produces a mechanically plausible curl at $\alpha = 1.5$: the lashes lift smoothly upward without sliding off the lid or stretching.

% \begin{figure}[t]
%     \centering
%     \includegraphics[width=0.7\columnwidth]{Figures/Exp/figure_application_eyelashCurvature.pdf}
%     \caption{\textbf{Eyelash curling.} Top: reconstructed eyelash geometry. Middle: arc-fitted geometry. Bottom: curvature-amplified result ($\times 1.5$).}
%     \label{fig:application_eyelashCurling}
% \end{figure}

% \begin{figure}[!t]
%     \centering
%     \includegraphics[width=\columnwidth]{Figures/Exp/figure_application_physSimul.pdf}
%     \caption{\textbf{Beard wind simulation.} Left: reconstructed strand geometry. Right: dynamically simulated result under a wind force, rendered in Blender Cycles. The simulation is performed in pure numpy using a damped Verlet integrator with Follow-The-Leader inextensibility constraints.}
%     \label{fig:application_simulation}
% \end{figure}

% \clearpage

\subsection{Physic Simulation}

\begin{figure*}[!h]
    \centering
    \vspace{3mm}
    \begin{subfigure}[t]{0.42\linewidth}
        \centering
        \includegraphics[width=\linewidth]{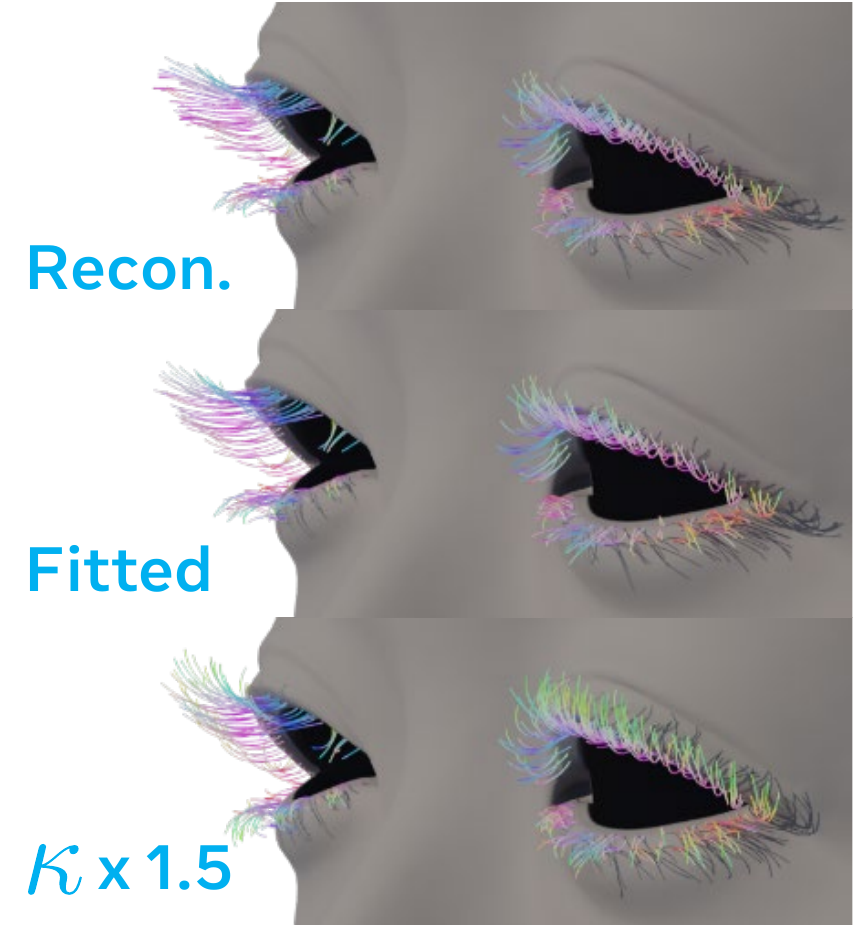}
        \caption{Eyelash curling.}
        \label{fig:application_eyelashCurling}
    \end{subfigure}
    \hfill
    \begin{subfigure}[t]{0.56\linewidth}
        \centering
        \includegraphics[width=\linewidth]{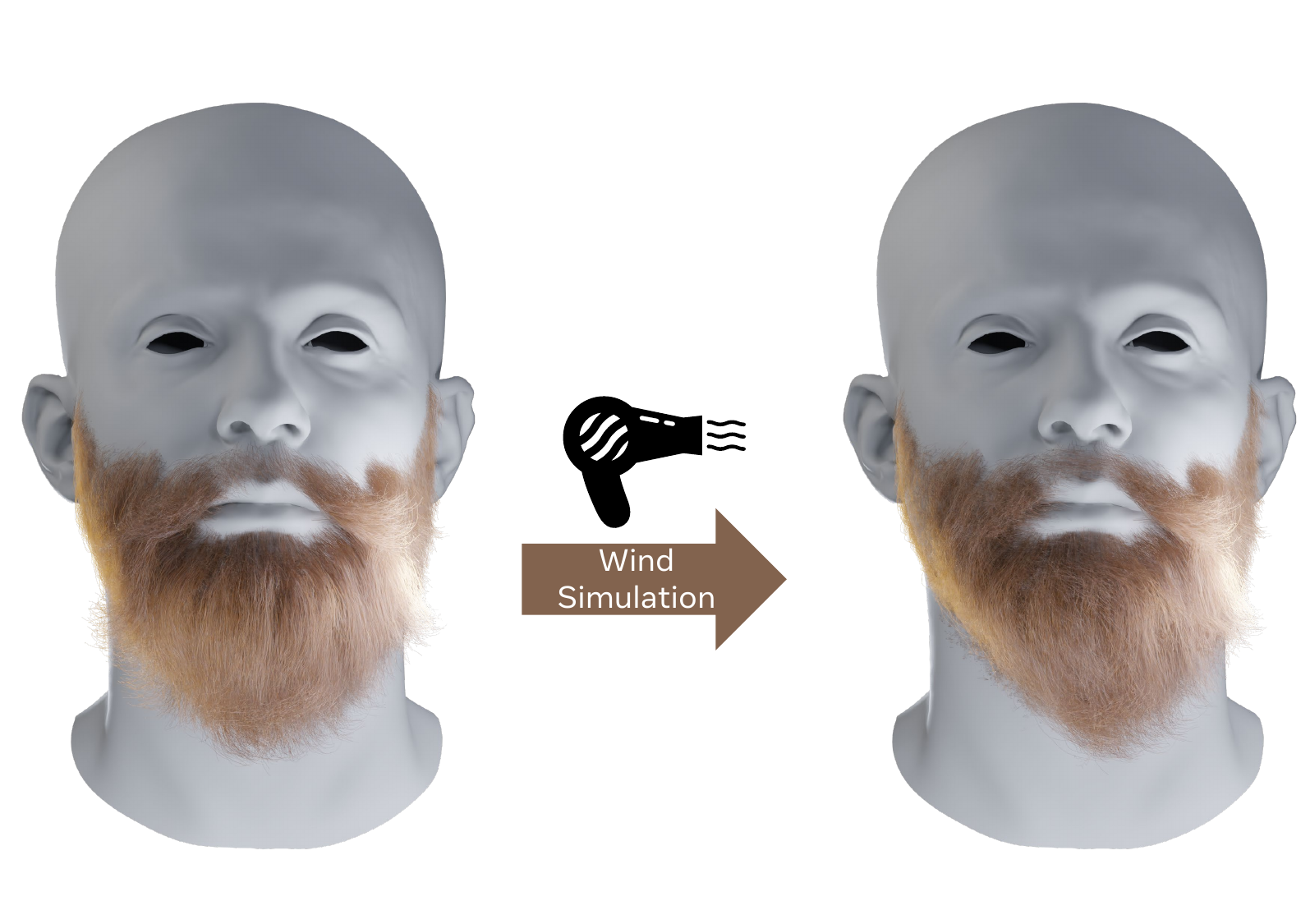}
        \caption{Beard wind simulation.}
        \label{fig:application_simulation}
    \end{subfigure}
    \caption{\textbf{Downstream applications.}
    \textbf{(a)} Eyelash curling. Top: reconstructed eyelash geometry. Middle: arc-fitted geometry. Bottom: curvature-amplified result ($\times 1.5$).
    \textbf{(b)} Beard wind simulation. Left: reconstructed strand geometry. Right: dynamically simulated result under a wind force, rendered in Blender Cycles. The simulation is performed in pure NumPy using a damped Verlet integrator with Follow-The-Leader inextensibility constraints.}
    \label{fig:applications}
\end{figure*}

% \subsection{Wind Simulation}
\Cref{fig:application_simulation} shows the wind simulation result, demonstrating that our grounded strand assets respond naturally to external forces without stretching artifacts.
To verify that our recovered strands support downstream animation, we attach a lightweight wind simulator to the per-vertex point cloud. Dynamics run in pure numpy without rigging or DCC tools, baking $\sim$34k strands (2.2M vertices) in a few seconds on a single CPU.

Each vertex is advanced with a damped Verlet step — more energy-conservative than explicit Euler integration — under three forces: gravity, a linear rest-pose spring, and a tip-emphasized wind force
\begin{equation}
\mathbf{F}^{\mathrm{wind}}_i = A\,\phi^3(i)\,g_i(t)\,S_i\,O_i\,\mu_i\,\hat{\mathbf{w}},
\end{equation}
where $\phi^3$ emphasizes tips, $g_i(t)$ is a spatiotemporal Perlin gust, $S_i$ a hemisphere shield, $O_i$ a head-mesh occlusion factor, and $\mu_i$ an inverse-length boost giving short stubble up to $2.5\times$ the force of long fibers. A Follow-The-Leader pass per strand enforces inextensibility, and root vertices are blended toward rest pose over the first 40\% of each strand to prevent unphysical follicle motion. The resulting clip shows windward beard sweep and pronounced motion of under-chin fibers, confirming that our reconstructed strands are free of stretching artifacts under physically-motivated forcing.

\subsection{Beard Cutting}
\begin{figure}[!h]
    \centering
    \includegraphics[width=1.0\textwidth]{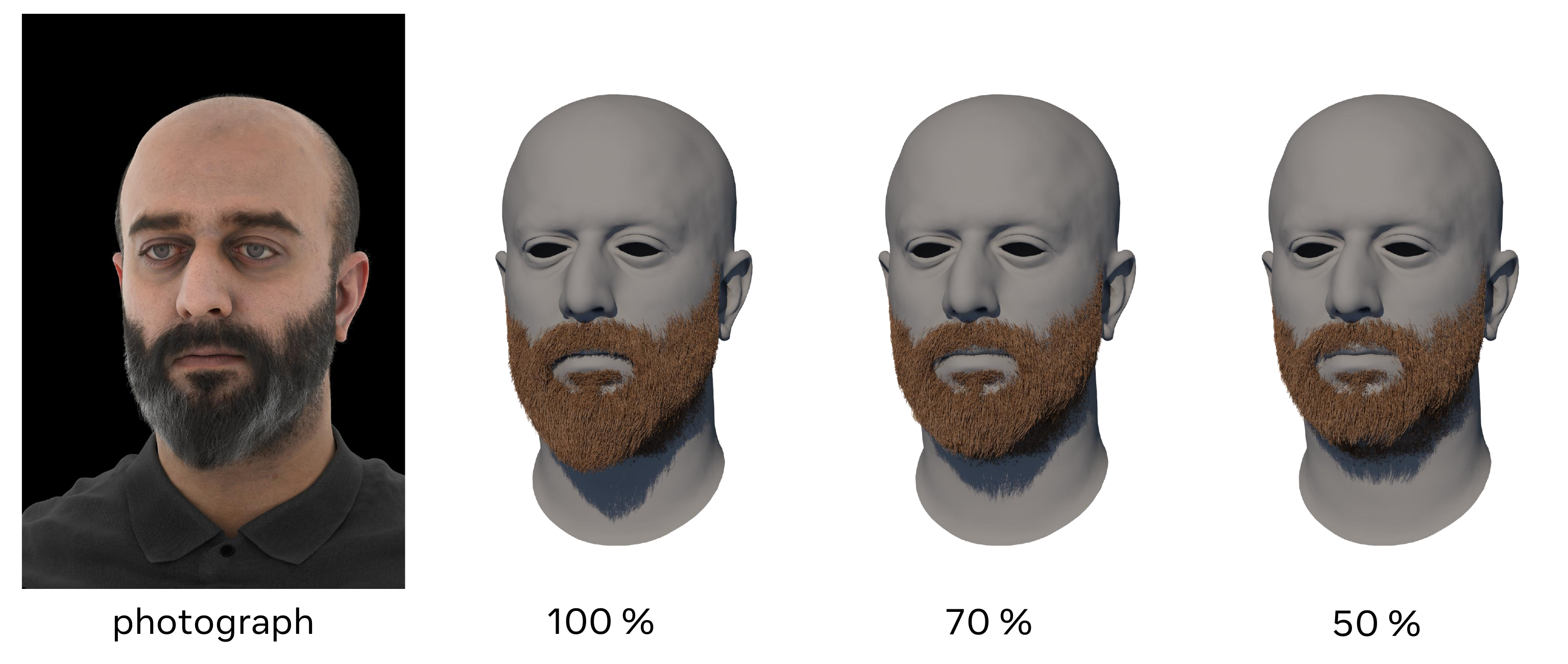}
    \caption{\textbf{Beard cutting.} Because each strand is grounded with a well-defined root-to-tip orientation, length-based cutting can be applied seamlessly by trimming strands at a fixed percentage of their total curve length from the root. This produces natural, style-consistent results across varying beard lengths (100\%, 70\%, 50\%), without any post-processing or manual intervention.}
    \label{fig:application_cutting}
\end{figure}

% \subsection{Beard Cutting}
Because each strand is grounded with a well-defined root-to-tip orientation, length-based cutting is straightforward: strands are trimmed at a fixed percentage of their total arc length from the root. No re-fitting, retraining, or manual intervention is required. As shown in~\Cref{fig:application_cutting}, this produces natural, style-consistent results across varying beard lengths, demonstrating that physically plausible grounding is a prerequisite for seamless grooming operations.

\subsection{Physical-based Lighting}

We render strands in Blender 4.2 with the Cycles path tracer using the Principled Hair BSDF~\cite{chiang2015practical}, which integrates R, TT, and TRT scattering lobes. Strand geometry is represented as cubic Bézier curves with a per-strand bevel radius of $2\,\mu m$, approximating an average human hair fiber. Material is parametrized via (melanin, melanin-redness, longitudinal/radial roughness) — e.g.\ chestnut $(0.65, 0.7, 0.40)$ and jet-black $(1.0, 0.5, 0.30)$ — or by directly overriding the absorption color for stylized appearances such as green. We support three lighting setups: (i) a Nishita sky model, (ii) the original studio acquisition rig, and (iii) a single area light orbiting the head, which decouples view-dependent strand effects from camera motion and provides the cleanest probe for material differences.

\begin{figure}[!h]
    \centering
    \includegraphics[width=\columnwidth]{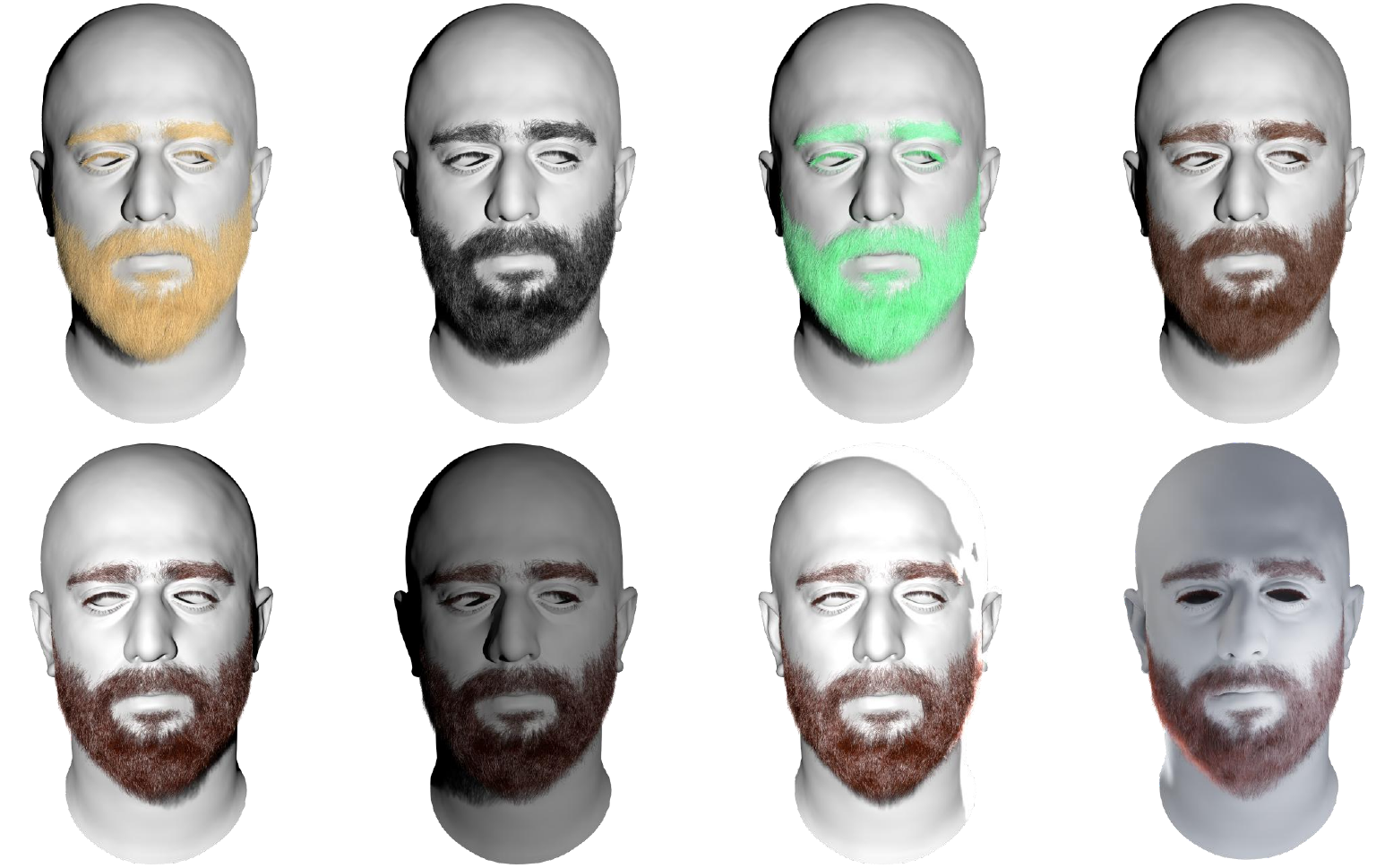}
    \caption{\textbf{Physics-based relighting and material editing.} Our grounded strand asset supports physically based rendering via the Chiang hair BSDF~\cite{chiang2015practical}. Top row: material editing with varying hair color (blonde, black, green, brown) under fixed lighting. Bottom row: the same geometry rendered under different lighting conditions with fixed material, demonstrating consistent strand-level response to illumination changes.}
    \label{fig:application_relighting}
\end{figure}

As shown in~\Cref{fig:application_relighting}, the top row demonstrates material editing under fixed lighting, and the bottom row shows relighting with fixed material, confirming that our reconstructed strands respond correctly to illumination changes under the BSDF model.

\clearpage

\section{Additional Ablation Studies}

\subsection{Sensitivity to Input Matting Quality}

For our experiments, we obtain facial hair masks using MatteAnything~\cite{yao2024matteanything} in an interactive setting, followed by manual cleanup of noisy regions in Photoshop to remove spurious background artifacts. As shown in~\Cref{fig:ablation_matting}, the quality of the input mask directly affects the level of detail in the reconstructed strand boundary.

While recent large-scale diffusion-based matting models have shown impressive general performance, they still struggle with facial hair and other regions exhibiting subsurface ambiguities, such as self-shadowing and ambient occlusion, where the boundary between skin and hair is inherently ill-defined. We believe that a dedicated facial-hair matting model, trained on purpose-built synthetic data with accurate ambient-occlusion and self-shadowing cues, would substantially improve reconstruction quality in these ambiguous boundary regions. We leave this as a promising direction for future work.

\subsection{Robustness on Different Albedos}
\begin{figure*}[!h]
    \centering
    \vspace{3mm}
    \begin{subfigure}[t]{0.49\linewidth}
        \centering
        \includegraphics[width=\linewidth]{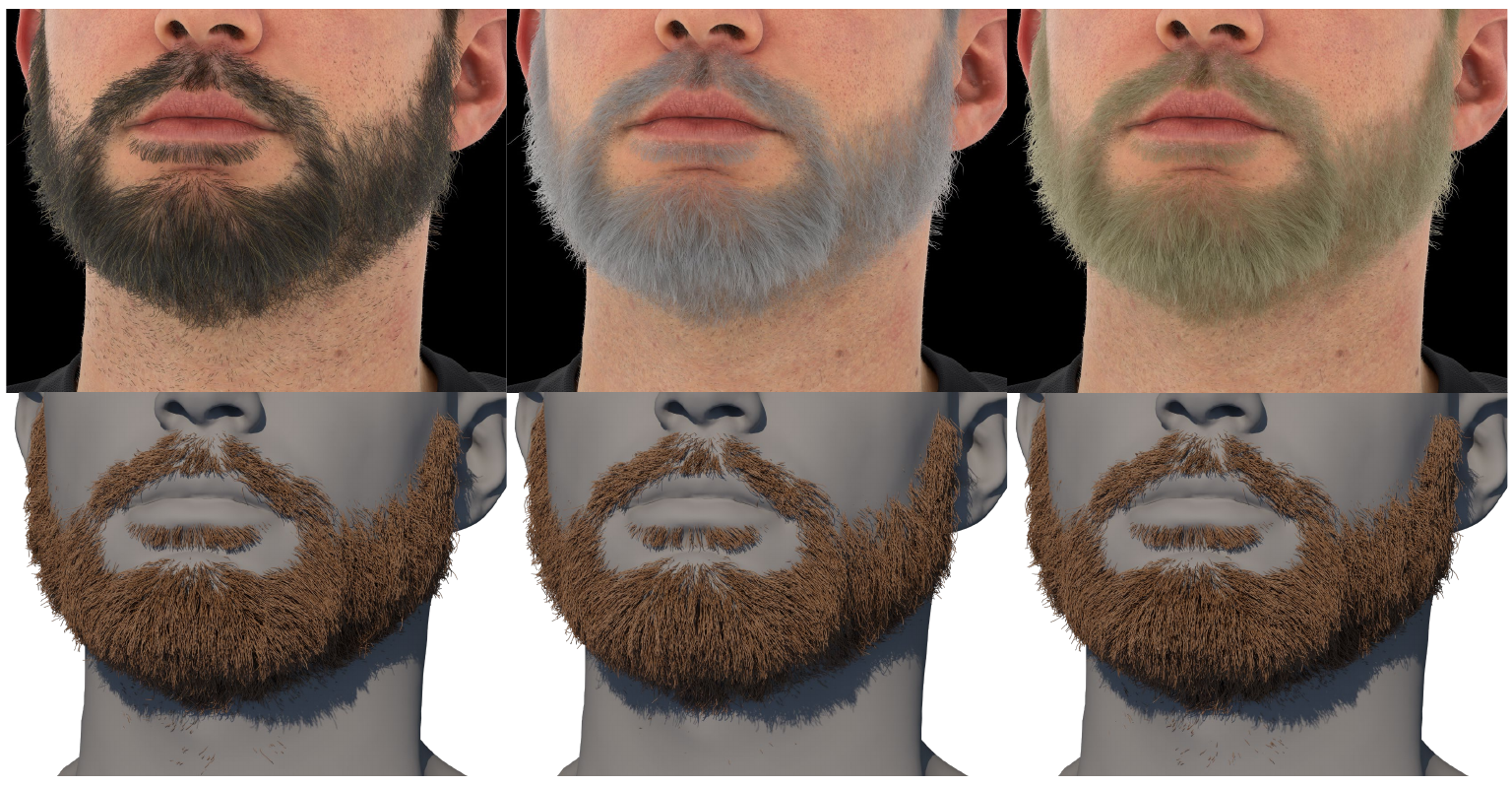}
        \caption{Robustness on different albedos.}
        \label{fig:robustness_color}
    \end{subfigure}
    \hfill
    \begin{subfigure}[t]{0.49\linewidth}
        \centering
        \includegraphics[width=\linewidth]{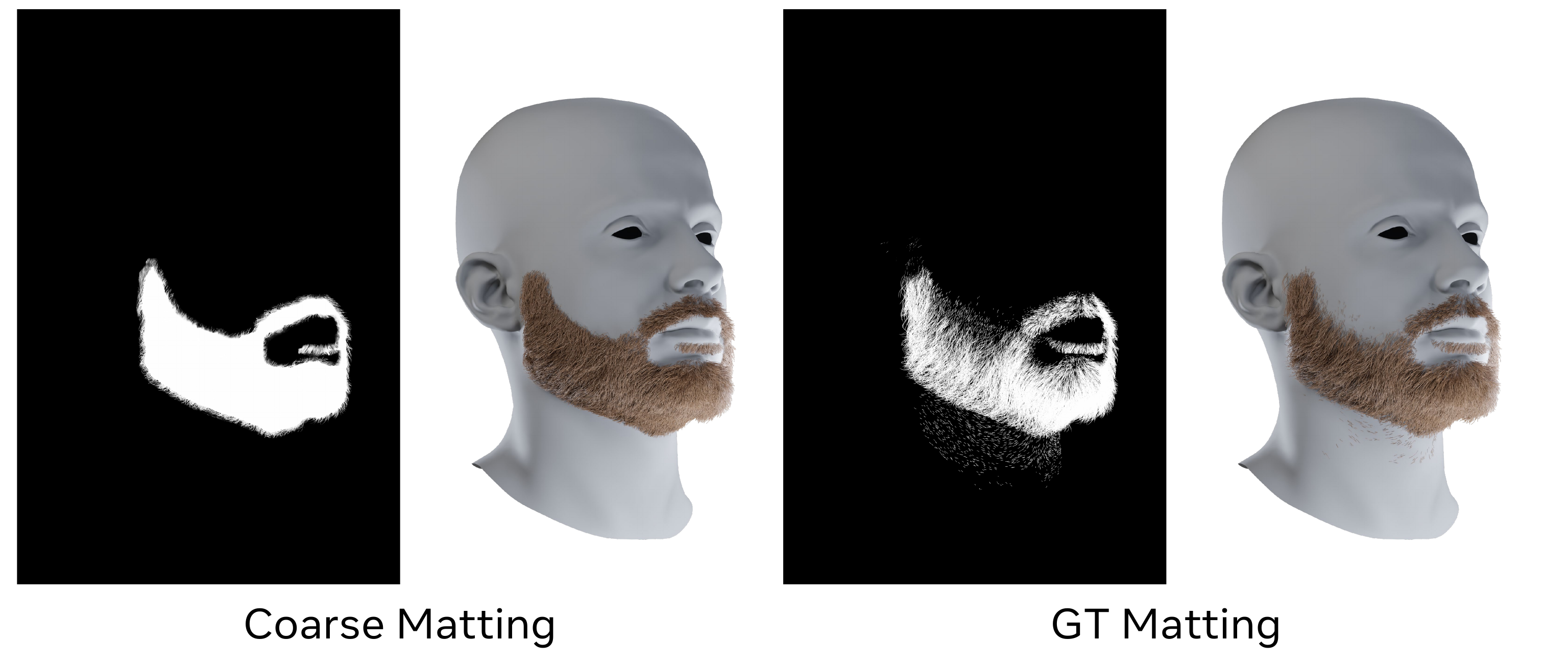}
        \caption{Effect of input matting quality.}
        \label{fig:ablation_matting}
    \end{subfigure}
    \caption{\textbf{Robustness and sensitivity analysis.}
    \textbf{(a)} Our method successfully reconstructs facial hair geometry across a range of hair colors, including dark, grey, and dyed variants (top row). The recovered strand geometry is consistent across albedos, demonstrating that our reconstruction is not biased toward specific hair colors. The bottom row shows the same geometry rendered with uniformly applied BSDF parameters.
    \textbf{(b)} Our method is sensitive to the quality of the input hair mask. With coarse matting (left), the overall strand distribution and growth trend are faithfully recovered, but fine boundary details such as wispy strand tips are lost, resulting in a harder silhouette. GT matting (right) preserves these fine-grained details, enabling more accurate reconstruction of the natural beard periphery.}
    \label{fig:robustness_matting}
\end{figure*}

A practical concern for any reconstruction method is whether it degrades on dark or unusually colored hair, as low-albedo regions exhibit weaker photometric gradients and reduced orientation contrast. As shown in~\Cref{fig:robustness_color}, our method produces geometrically consistent strand reconstructions across dark, grey, and dyed green hair colors. We attribute this robustness to our depth-gated formulation, which constrains the optimization to valid hair support regardless of albedo, and to the orientation supervision which provides a signal independent of absolute color.

\section{Additional Comparison}
\subsection{Numerical Comparison}
Table~\ref{tab:quantitative} reveals a complementary pattern across facial hair types. For eyebrows, LPMVS achieves competitive or superior precision and F1 at tight thresholds, where the short, straight, and non-occluded strand geometry favors line-based stereo matching. However, our method consistently outperforms LPMVS in Strand Consistency across all regions and thresholds, indicating that LPMVS tends to fragment strands into short disconnected segments even when point-level coverage is high — a limitation that becomes increasingly pronounced for curved and occluded structures such as eyelashes and beards. For beard reconstruction, where dense crossings and heavy occlusion cause LPMVS line search to fail and recall to collapse, our method achieves substantially higher F1 and SC. HairGS consistently underperforms across all regions, particularly for eyebrows where its volumetric formulation lacks the spatial resolution to reconstruct thin, localized structures.

\begin{table*}[t]
\centering
\vspace{3mm}
\caption{\textbf{Quantitative comparison of facial hair reconstruction.} We report F1 and Strand Consistency (SC$^\dagger$) at multiple distance/angle thresholds, averaged over $n$ subjects per region. \textbf{Bold} indicates best per metric. Per-metric breakdown including Precision and Recall is provided in the supplementary material.}
\label{tab:quantitative}
\setlength{\tabcolsep}{3pt}
\renewcommand{\arraystretch}{1.1}
\small
\resizebox{\linewidth}{!}{%
\begin{tabular}{l *{6}{c} *{6}{c} *{6}{c}}
\toprule
& \multicolumn{6}{c}{\textit{Eyebrow} ($n{=}5$)} & \multicolumn{6}{c}{\textit{Eyelash} ($n{=}4$)} & \multicolumn{6}{c}{\textit{Beard} ($n{=}3$)} \\
\cmidrule(lr){2-7} \cmidrule(lr){8-13} \cmidrule(lr){14-19}
& \multicolumn{2}{c}{$0.1\,\text{mm},10°$} & \multicolumn{2}{c}{$0.2\,\text{mm},20°$} & \multicolumn{2}{c}{$0.5\,\text{mm},30°$}
& \multicolumn{2}{c}{$0.1\,\text{mm},10°$} & \multicolumn{2}{c}{$0.2\,\text{mm},20°$} & \multicolumn{2}{c}{$0.5\,\text{mm},30°$}
& \multicolumn{2}{c}{$0.5\,\text{mm},20°$} & \multicolumn{2}{c}{$1.0\,\text{mm},30°$} & \multicolumn{2}{c}{$1.0\,\text{mm},90°$} \\
\cmidrule(lr){2-3} \cmidrule(lr){4-5} \cmidrule(lr){6-7} \cmidrule(lr){8-9} \cmidrule(lr){10-11} \cmidrule(lr){12-13} \cmidrule(lr){14-15} \cmidrule(lr){16-17} \cmidrule(lr){18-19}
Method & F1 & SC & F1 & SC & F1 & SC & F1 & SC & F1 & SC & F1 & SC & F1 & SC & F1 & SC & F1 & SC \\
\midrule
LPMVS  & 0.583 & 0.480 & \textbf{0.807} & 0.803 & \textbf{0.956} & 0.968 & \textbf{0.748} & 0.514 & 0.966 & 0.849 & \textbf{0.996} & 0.977 & 0.599 & 0.374 & 0.871 & 0.648 & 0.952 & 0.887 \\
HairGS & 0.010 & 0.064 & 0.055 & 0.408 & 0.242 & 0.833 & 0.089 & 0.072 & 0.297 & 0.349 & 0.493 & 0.677 & 0.518 & 0.431 & 0.806 & 0.779 & 0.866 & 0.966 \\
Ours   & \textbf{0.622} & \textbf{0.541} & 0.806 & \textbf{0.854} & 0.953 & \textbf{0.985} & 0.729 & \textbf{0.641} & \textbf{0.972} & \textbf{0.966} & \textbf{0.996} & \textbf{0.996} & \textbf{0.922} & \textbf{0.630} & \textbf{0.992} & \textbf{0.867} & \textbf{0.996} & \textbf{0.991} \\
\bottomrule
\end{tabular}%
}
\vspace{2pt}
\begin{minipage}{\linewidth}
\footnotesize $^\dagger$SC (Strand Consistency): fraction of each GT strand covered by a single best-matching predicted strand, averaged over all GT strands. Unlike F1, SC penalizes fragmented predictions even when point-level coverage is high.
\end{minipage}
\end{table*}

\clearpage
\newpage
\bibliographystyle{assets/plainnat}
\bibliography{paper}

@String(TOG= {ACM Trans. Graph.})

@String(ICLR = {Int. Conf. Learn. Represent.})

@String(TOG   = {ACM TOG})

@String(ICLR  = {ICLR})

@article {hu2015uschair,
    	Author = {Liwen Hu and Chongyang Ma and Linjie Luo and Hao Li},
	Title = {Single-View Hair Modeling Using A Hairstyle Database},
  	Journal = {ACM Transactions on Graphics (Proceedings SIGGRAPH 2015)},
	Volume = {34},
	Number = {4},
  	Year = {2015},
	Month = {July},
 	Publisher = {ACM},
 	Paddress = {New York, NY, USA}
}

@inproceedings{gaussianhaircut,
  title={Human hair reconstruction with strand-aligned 3d gaussians},
  author={Zakharov, Egor and Sklyarova, Vanessa and Black, Michael and Nam, Giljoo and Thies, Justus and Hilliges, Otmar},
  booktitle={European Conference on Computer Vision},
  pages={409--425},
  year={2024},
  organization={Springer}
}

@article{gaussianhair,
  title={Gaussianhair: Hair modeling and rendering with light-aware gaussians},
  author={Luo, Haimin and Ouyang, Min and Zhao, Zijun and Jiang, Suyi and Zhang, Longwen and Zhang, Qixuan and Yang, Wei and Xu, Lan and Yu, Jingyi},
  journal={arXiv preprint arXiv:2402.10483},
  year={2024}
}

@article{groomcap,
  title={Groomcap: High-fidelity prior-free hair capture},
  author={Zhou, Yuxiao and Chai, Menglei and Wang, Daoye and Winberg, Sebastian and Wood, Erroll and Sarkar, Kripasindhu and Gross, Markus and Beeler, Thabo},
  journal={ACM Transactions on Graphics (TOG)},
  volume={43},
  number={6},
  pages={1--15},
  year={2024},
  publisher={ACM New York, NY, USA}
}

@article{groomgen,
  title={Groomgen: A high-quality generative hair model using hierarchical latent representations},
  author={Zhou, Yuxiao and Chai, Menglei and Pepe, Alessandro and Gross, Markus and Beeler, Thabo},
  journal={ACM Transactions on Graphics (TOG)},
  volume={42},
  number={6},
  pages={1--16},
  year={2023},
  publisher={ACM New York, NY, USA}
}

@inproceedings{lpmvs,
  title={Strand-accurate multi-view hair capture},
  author={Nam, Giljoo and Wu, Chenglei and Kim, Min H and Sheikh, Yaser},
  booktitle={Proceedings of the IEEE/CVF Conference on Computer Vision and Pattern Recognition},
  pages={155--164},
  year={2019}
}

@article{pan2025hairgs,
  title={HairGS: Hair Strand Reconstruction based on 3D Gaussian Splatting},
  author={Pan, Yimin and Niessner, Matthias and Kirschstein, Tobias},
  journal={arXiv preprint arXiv:2509.07774},
  year={2025}
}

@article{ct2hair,
  title={Ct2hair: High-fidelity 3d hair modeling using computed tomography},
  author={Shen, Yuefan and Saito, Shunsuke and Wang, Ziyan and Maury, Olivier and Wu, Chenglei and Hodgins, Jessica and Zheng, Youyi and Nam, Giljoo},
  journal={ACM Transactions on Graphics (TOG)},
  volume={42},
  number={4},
  pages={1--13},
  year={2023},
  publisher={ACM New York, NY, USA}
}

@inproceedings{groomlight,
  title={GroomLight: Hybrid Inverse Rendering for Relightable Human Hair Appearance Modeling},
  author={Zheng, Yang and Chai, Menglei and Vicini, Delio and Zhou, Yuxiao and Xu, Yinghao and Guibas, Leonidas and Wetzstein, Gordon and Beeler, Thabo},
  booktitle={Proceedings of the Computer Vision and Pattern Recognition Conference},
  pages={16040--16050},
  year={2025}
}

@article{hairinverserendering,
  title={Human hair inverse rendering using multi-view photometric data},
  author={Sun, Tiancheng and Nam, Giljoo and Aliaga, Carlos and Hery, Christophe and Ramamoorthi, Ravi},
  year={2021},
  publisher={The Eurographics Association}
}

@article{beelerfacialhair,
  title={Coupled 3D reconstruction of sparse facial hair and skin},
  author={Beeler, Thabo and Bickel, Bernd and Noris, Gioacchino and Beardsley, Paul and Marschner, Steve and Sumner, Robert W and Gross, Markus},
  journal={ACM Transactions on Graphics (ToG)},
  volume={31},
  number={4},
  pages={1--10},
  year={2012},
  publisher={ACM New York, NY, USA}
}

@inproceedings{grabli02,
  title={Image-based hair capture by inverse lighting},
  author={Grabli, St{\'e}phane and Sillion, Fran{\c{c}}ois X and Marschner, Stephen R and Lengyel, Jerome E},
  booktitle={Proceedings of Graphics Interface (GI)},
  pages={51--58},
  year={2002}
}

@article{paris04,
  title={Capture of hair geometry from multiple images},
  author={Paris, Sylvain and Briceno, Hector M and Sillion, Fran{\c{c}}ois X},
  journal={ACM transactions on graphics (TOG)},
  volume={23},
  number={3},
  pages={712--719},
  year={2004},
  publisher={ACM New York, NY, USA}
}

@article{paris08,
  title={Hair photobooth: geometric and photometric acquisition of real hairstyles.},
  author={Paris, Sylvain and Chang, Will and Kozhushnyan, Oleg I and Jarosz, Wojciech and Matusik, Wojciech and Zwicker, Matthias and Durand, Fr{\'e}do},
  journal={ACM Trans. Graph.},
  volume={27},
  number={3},
  pages={30},
  year={2008}
}

@inproceedings{kerbiriou2024eyelash,
  title={3D Reconstruction and Semantic Modeling of Eyelashes},
  author={Kerbiriou, Glenn and Avril, Quentin and Marchal, Maud},
  booktitle={Computer Graphics Forum},
  volume={43},
  number={2},
  pages={e15040},
  year={2024},
  organization={Wiley Online Library}
}

@inproceedings{erank,
  title={The effective rank: A measure of effective dimensionality},
  author={Roy, Olivier and Vetterli, Martin},
  booktitle={2007 15th European signal processing conference},
  pages={606--610},
  year={2007},
  organization={IEEE}
}

@article{erankGaussian,
  title={Effective rank analysis and regularization for enhanced 3d gaussian splatting},
  author={Hyung, Junha and Hong, Susung and Hwang, Sungwon and Lee, Jaeseong and Choo, Jaegul and Kim, Jin-Hwa},
  journal={Advances in Neural Information Processing Systems},
  volume={37},
  pages={110412--110435},
  year={2024}
}

@article{winberg2022facial,
  title={Facial hair tracking for high fidelity performance capture},
  author={Winberg, Sebastian and Zoss, Gaspard and Chandran, Prashanth and Gotardo, Paulo and Bradley, Derek},
  journal={ACM Transactions on Graphics (TOG)},
  volume={41},
  number={4},
  pages={1--12},
  year={2022},
  publisher={ACM New York, NY, USA}
}

@article{li2024strandfacialhair,
  title={Strand-accurate multi-view facial hair reconstruction and tracking},
  author={Li, Hanchao and Liu, Xinguo},
  journal={The Visual Computer},
  volume={40},
  number={7},
  pages={4713--4724},
  year={2024},
  publisher={Springer}
}

@article{li2023ems,
  title={EMS: 3D Eyebrow Modeling from Single-view Images},
  author={Li, Chenghong and Jin, Leyang and Zheng, Yujian and Yu, Yizhou and Han, Xiaoguang},
  journal={ACM Transactions on Graphics (TOG)},
  volume={42},
  number={6},
  pages={1--19},
  year={2023},
  publisher={ACM New York, NY, USA}
}

@article{maurer2016malebeard,
  title={The male beard hair and facial skin--challenges for shaving},
  author={Maurer, Marcus and Rietzler, Miriam and Burghardt, Renata and Siebenhaar, Frank},
  journal={International journal of cosmetic science},
  volume={38},
  pages={3--9},
  year={2016},
  publisher={Wiley Online Library}
}

@article{leao2017beard,
  title={Beard transplantation},
  author={LE{\~A}O, CARLOS EDUARDO GUIMAR{\~A}ES},
  journal={Revista Brasileira de Cirurgia Pl{\'a}stica},
  volume={32},
  pages={314--320},
  year={2017},
  publisher={SciELO Brasil}
}

@article{kerbl20233dgs,
  title={3d gaussian splatting for real-time radiance field rendering.},
  author={Kerbl, Bernhard and Kopanas, Georgios and Leimk{\"u}hler, Thomas and Drettakis, George and others},
  journal={ACM Trans. Graph.},
  volume={42},
  number={4},
  pages={139--1},
  year={2023}
}

@inproceedings{wu2024monohair,
  title={Monohair: High-fidelity hair modeling from a monocular video},
  author={Wu, Keyu and Yang, Lingchen and Kuang, Zhiyi and Feng, Yao and Han, Xutao and Shen, Yuefan and Fu, Hongbo and Zhou, Kun and Zheng, Youyi},
  booktitle={Proceedings of the IEEE/CVF Conference on Computer Vision and Pattern Recognition},
  pages={24164--24173},
  year={2024}
}

@inproceedings{wu2022neuralhdhair,
  title={Neuralhdhair: Automatic high-fidelity hair modeling from a single image using implicit neural representations},
  author={Wu, Keyu and Ye, Yifan and Yang, Lingchen and Fu, Hongbo and Zhou, Kun and Zheng, Youyi},
  booktitle={Proceedings of the IEEE/CVF Conference on Computer Vision and Pattern Recognition},
  pages={1526--1535},
  year={2022}
}

@inproceedings{rosu2025difflocks,
  title={Difflocks: Generating 3d hair from a single image using diffusion models},
  author={Rosu, Radu Alexandru and Wu, Keyu and Feng, Yao and Zheng, Youyi and Black, Michael J},
  booktitle={Proceedings of the Computer Vision and Pattern Recognition Conference},
  pages={10847--10857},
  year={2025}
}

@inproceedings{zheng2023hairstep,
  title={Hairstep: Transfer synthetic to real using strand and depth maps for single-view 3d hair modeling},
  author={Zheng, Yujian and Jin, Zirong and Li, Moran and Huang, Haibin and Ma, Chongyang and Cui, Shuguang and Han, Xiaoguang},
  booktitle={Proceedings of the IEEE/CVF Conference on Computer Vision and Pattern Recognition},
  pages={12726--12735},
  year={2023}
}

@inproceedings{sklyarova2023neuralhaircut,
  title={Neural haircut: Prior-guided strand-based hair reconstruction},
  author={Sklyarova, Vanessa and Chelishev, Jenya and Dogaru, Andreea and Medvedev, Igor and Lempitsky, Victor and Zakharov, Egor},
  booktitle={Proceedings of the IEEE/CVF International Conference on Computer Vision},
  pages={19762--19773},
  year={2023}
}

@inproceedings{rosu2022neuralstrands,
  title={Neural strands: Learning hair geometry and appearance from multi-view images},
  author={Rosu, Radu Alexandru and Saito, Shunsuke and Wang, Ziyan and Wu, Chenglei and Behnke, Sven and Nam, Giljoo},
  booktitle={European Conference on Computer Vision},
  pages={73--89},
  year={2022},
  organization={Springer}
}

@article{lombardi2019neuralvolume,
  title={Neural volumes: Learning dynamic renderable volumes from images},
  author={Lombardi, Stephen and Simon, Tomas and Saragih, Jason and Schwartz, Gabriel and Lehrmann, Andreas and Sheikh, Yaser},
  journal={arXiv preprint arXiv:1906.07751},
  year={2019}
}

@article{xiao2021eyelashnet,
  title={Eyelashnet: A dataset and a baseline method for eyelash matting},
  author={Xiao, Qinjie and Zhang, Hanyuan and Zhang, Zhaorui and Wu, Yiqian and Wang, Luyuan and Jin, Xiaogang and Jiang, Xinwei and Yang, Yong-Liang and Shao, Tianjia and Zhou, Kun},
  journal={ACM Transactions on Graphics (TOG)},
  volume={40},
  number={6},
  pages={1--17},
  year={2021},
  publisher={ACM New York, NY, USA}
}

@inproceedings{qian2024gaussianavatars,
  title={Gaussianavatars: Photorealistic head avatars with rigged 3d gaussians},
  author={Qian, Shenhan and Kirschstein, Tobias and Schoneveld, Liam and Davoli, Davide and Giebenhain, Simon and Nie{\ss}ner, Matthias},
  booktitle={Proceedings of the IEEE/CVF Conference on Computer Vision and Pattern Recognition},
  pages={20299--20309},
  year={2024}
}

@article{lombardi2021mvp,
  title={Mixture of volumetric primitives for efficient neural rendering},
  author={Lombardi, Stephen and Simon, Tomas and Schwartz, Gabriel and Zollhoefer, Michael and Sheikh, Yaser and Saragih, Jason},
  journal={ACM Transactions on Graphics (ToG)},
  volume={40},
  number={4},
  pages={1--13},
  year={2021},
  publisher={ACM New York, NY, USA}
}

@inproceedings{lee2025surfhead,
  title={SURFHEAD: AFFINE RIG BLENDING FOR GEOMETRICALLY ACCURATE 2D GAUSSIAN SURFEL HEAD AVATARS},
  author={Lee, Jaeseong and Kang, Taewoong and B{\"u}hler, Marcel C and Kim, Min-Jung and Hwang, Sungwon and Hyung, Junha and Jang, Hyojin and Choo, Jaegul},
  booktitle={13th International Conference on Learning Representations, ICLR 2025},
  pages={2877--2901},
  year={2025},
  organization={International Conference on Learning Representations, ICLR}
}

@inproceedings{lee2025texavatars,
  title={TexAvatars: Hybrid Texel-3D Representations for Stable Rigging of Photorealistic Gaussian Head Avatars},
  author={Lee, Jaeseong and Ahn, Junyeong and Kang, Taewoong and Choo, Jaegul},
  booktitle={Thirteenth International Conference on 3D Vision},
  year={2025}
}

@inproceedings{zheng2022imavatar,
  title={Im avatar: Implicit morphable head avatars from videos},
  author={Zheng, Yufeng and Abrevaya, Victoria Fern{\'a}ndez and B{\"u}hler, Marcel C and Chen, Xu and Black, Michael J and Hilliges, Otmar},
  booktitle={Proceedings of the IEEE/CVF conference on computer vision and pattern recognition},
  pages={13545--13555},
  year={2022}
}

@inproceedings{zheng2023pointavatar,
  title={Pointavatar: Deformable point-based head avatars from videos},
  author={Zheng, Yufeng and Yifan, Wang and Wetzstein, Gordon and Black, Michael J and Hilliges, Otmar},
  booktitle={Proceedings of the IEEE/CVF conference on computer vision and pattern recognition},
  pages={21057--21067},
  year={2023}
}

@inproceedings{giebenhain2023nphm,
  title={Learning neural parametric head models},
  author={Giebenhain, Simon and Kirschstein, Tobias and Georgopoulos, Markos and R{\"u}nz, Martin and Agapito, Lourdes and Nie{\ss}ner, Matthias},
  booktitle={Proceedings of the IEEE/CVF Conference on Computer Vision and Pattern Recognition},
  pages={21003--21012},
  year={2023}
}

@inproceedings{gafni2021nerface,
  title={Dynamic neural radiance fields for monocular 4d facial avatar reconstruction},
  author={Gafni, Guy and Thies, Justus and Zollhofer, Michael and Nie{\ss}ner, Matthias},
  booktitle={Proceedings of the IEEE/CVF conference on computer vision and pattern recognition},
  pages={8649--8658},
  year={2021}
}

@article{bharadwaj2023flare,
  title={Flare: Fast learning of animatable and relightable mesh avatars},
  author={Bharadwaj, Shrisha and Zheng, Yufeng and Hilliges, Otmar and Black, Michael J and Fernandez-Abrevaya, Victoria},
  journal={arXiv preprint arXiv:2310.17519},
  year={2023}
}

@article{feng2021deca,
  title={Learning an animatable detailed 3D face model from in-the-wild images},
  author={Feng, Yao and Feng, Haiwen and Black, Michael J and Bolkart, Timo},
  journal={ACM Transactions on Graphics (ToG)},
  volume={40},
  number={4},
  pages={1--13},
  year={2021},
  publisher={ACM New York, NY, USA}
}

@inproceedings{lattas2023fitme,
  title={Fitme: Deep photorealistic 3d morphable model avatars},
  author={Lattas, Alexandros and Moschoglou, Stylianos and Ploumpis, Stylianos and Gecer, Baris and Deng, Jiankang and Zafeiriou, Stefanos},
  booktitle={Proceedings of the IEEE/CVF Conference on Computer Vision and Pattern Recognition},
  pages={8629--8640},
  year={2023}
}

@inproceedings{papantoniou2023relightify,
  title={Relightify: Relightable 3d faces from a single image via diffusion models},
  author={Papantoniou, Foivos Paraperas and Lattas, Alexandros and Moschoglou, Stylianos and Zafeiriou, Stefanos},
  booktitle={Proceedings of the IEEE/CVF International Conference on Computer Vision},
  pages={8806--8817},
  year={2023}
}

@inproceedings{liu2025bioface,
  title={Controllable Biophysical Human Faces},
  author={Liu, Minghao and Grabli, Stephane and Speierer, S{\'e}bastien and Sarafianos, Nikolaos and Bode, Lukas and Chiang, Matt and Hery, Christophe and Davis, James and Aliaga, Carlos},
  booktitle={Computer Graphics Forum},
  volume={44},
  number={4},
  pages={e70170},
  year={2025},
  organization={Wiley Online Library}
}

@article{retsinas2024smirk,
  title={SMIRK: 3D Facial Expressions through Analysis-by-Neural-Synthesis},
  author={Retsinas, George and Filntisis, Panagiotis P and Danecek, Radek and Abrevaya, Victoria F and Roussos, Anastasios and Bolkart, Timo and Maragos, Petros},
  journal={arXiv preprint arXiv:2404.04104},
  year={2024}
}

@inproceedings{saito2024rgca,
  title={Relightable gaussian codec avatars},
  author={Saito, Shunsuke and Schwartz, Gabriel and Simon, Tomas and Li, Junxuan and Nam, Giljoo},
  booktitle={Proceedings of the IEEE/CVF conference on computer vision and pattern recognition},
  pages={130--141},
  year={2024}
}

@inproceedings{luo2012multi,
  title={Multi-view hair capture using orientation fields},
  author={Luo, Linjie and Li, Hao and Paris, Sylvain and Weise, Thibaut and Pauly, Mark and Rusinkiewicz, Szymon},
  booktitle={2012 IEEE Conference on Computer Vision and Pattern Recognition},
  pages={1490--1497},
  year={2012},
  organization={Ieee}
}

@inproceedings{luo2013wide,
  title={Wide-baseline hair capture using strand-based refinement},
  author={Luo, Linjie and Zhang, Cha and Zhang, Zhengyou and Rusinkiewicz, Szymon},
  booktitle={Proceedings of the IEEE Conference on Computer Vision and Pattern Recognition},
  pages={265--272},
  year={2013}
}

@article{herrera2012lighting,
  title={Lighting hair from the inside: A thermal approach to hair reconstruction},
  author={Herrera, Tomas Lay and Zinke, Arno and Weber, Andreas},
  journal={ACM Transactions on Graphics (TOG)},
  volume={31},
  number={6},
  pages={1--9},
  year={2012},
  publisher={ACM New York, NY, USA}
}

@inproceedings{takimoto2024drhair,
  title={Dr. hair: Reconstructing scalp-connected hair strands without pre-training via differentiable rendering of line segments},
  author={Takimoto, Yusuke and Takehara, Hikari and Sato, Hiroyuki and Zhu, Zihao and Zheng, Bo},
  booktitle={Proceedings of the IEEE/CVF Conference on Computer Vision and Pattern Recognition},
  pages={20601--20611},
  year={2024}
}

@inproceedings{Chang2025IPHG,
  author = {Chang, Wesley and Russell, Andrew L. and Grabli, Stephane and Chiang, Matt Jen-Yuan and Hery, Christophe and Roble, Doug and Ramamoorthi, Ravi and Li, Tzu-Mao and Maury, Olivier},
  title = {Transforming Unstructured Hair Strands into Procedural Hair Grooms},
  year = {2025},
  issue_date = {August 2025},
  publisher = {Association for Computing Machinery},
  address = {New York, NY, USA},
  volume = {44},
  number = {4},
  url = {https://doi.org/10.1145/3731168},
  doi = {10.1145/3731168},
  journal = {ACM Trans. Graph.},
  month = aug,
  articleno = {105},
  numpages = {20}
}

@inproceedings{sklyarova2025im2haircut,
  title={Im2Haircut: Single-view Strand-based Hair Reconstruction for Human Avatars},
  author={Sklyarova, Vanessa and Zakharov, Egor and Prinzler, Malte and Becherini, Giorgio and Black, Michael J and Thies, Justus},
  booktitle={Proceedings of the IEEE/CVF International Conference on Computer Vision},
  pages={10656--10665},
  year={2025}
}

@article{yao2024matteanything,
  title={Matte anything: Interactive natural image matting with segment anything model},
  author={Yao, Jingfeng and Wang, Xinggang and Ye, Lang and Liu, Wenyu},
  journal={Image and Vision Computing},
  pages={105067},
  year={2024},
  publisher={Elsevier}
}

@article{yariv2021volume,
  title={Volume rendering of neural implicit surfaces},
  author={Yariv, Lior and Gu, Jiatao and Kasten, Yoni and Lipman, Yaron},
  journal={Advances in neural information processing systems},
  volume={34},
  pages={4805--4815},
  year={2021}
}

@article{wang2021neus,
  title={Neus: Learning neural implicit surfaces by volume rendering for multi-view reconstruction},
  author={Wang, Peng and Liu, Lingjie and Liu, Yuan and Theobalt, Christian and Komura, Taku and Wang, Wenping},
  journal={arXiv preprint arXiv:2106.10689},
  year={2021}
}

@incollection{chiang2015practical,
  title={A practical and controllable hair and fur model for production path tracing},
  author={Chiang, Matt Jen-Yuan and Bitterli, Benedikt and Tappan, Chuck and Burley, Brent},
  booktitle={ACM SIGGRAPH 2015 Talks},
  pages={1--1},
  year={2015}
}

@software{jakob2022mitsuba3,
    title = {Mitsuba 3 renderer},
    author = {Wenzel Jakob and Sébastien Speierer and Nicolas Roussel and Merlin Nimier-David and Delio Vicini and Tizian Zeltner and Baptiste Nicolet and Miguel Crespo and Vincent Leroy and Ziyi Zhang},
    note = {https://mitsuba-renderer.org},
    version = {3.0.1},
    year = 2022,
}

@article{Laine2020diffrast,
  title   = {Modular Primitives for High-Performance Differentiable Rendering},
  author  = {Samuli Laine and Janne Hellsten and Tero Karras and Yeongho Seol and Jaakko Lehtinen and Timo Aila},
  journal = {ACM Transactions on Graphics},
  year    = {2020},
  volume  = {39},
  number  = {6}
}

\clearpage
\newpage
% \beginappendix

% \section{First appendix}

\end{document}